\documentclass[preprint, amsmath, amssymb, aps, groupedaddress]{revtex4-1}
\usepackage{graphicx, subfigure}
\usepackage{dcolumn}
\usepackage{bm}

\usepackage{indentfirst}
\usepackage{array, threeparttable}
\usepackage{multirow}
\usepackage{mathrsfs}
\usepackage{color}

\begin{document}
	
	
\title{A third order gas-kinetic scheme for unstructured grid}
	
	
\author{Ji Li}
\email[]{leejearl@mail.nwpu.edu.cn}
\author{Chengwen Zhong}
\email[Corresponding author: ]{zhongcw@nwpu.edu.cn}
\author{Congshan Zhuo}%
\email{zhuocs@nwpu.edu.cn}

\affiliation{National Key Laboratory of Science and Technology on Aerodynamic Design and Research, Northwestern Polytechnical University, Xi'an, Shaanxi 710072, China.}
	
\date{\today}
	
\begin{abstract}
In our study, a compact third order gas-kinetic scheme is constructed for unstructured grid which is combined the compact least-square reconstruction (CLS) method. The CLS method can achieve arbitrary high order compact reconstruction using the stencil from the whole computational domain implicitly. A large sparse linear system resulted from the CLS reconstruction method is solved by applying the generalized minimal residual algorithm (GMRES), and the Reverse-Cuthill-McKee (RCM) algorithm and the incomplete lower-upper (ILU) factorization method are implemented to accelerate the convergence of the iterative method. Different from the traditional flux solver, the BGK based gas-kinetic scheme is in the nature of spatial and temporal accuracy. Applying the second order expansion to the distribution function, the third order flux solver can be obtained directly. The accuracy of present method is validated by several numerical cases such as the advection of density perturbation problem, Sod shock wave problem, Lax shock tube test case, Shu-Osher problem, shock-vortex interaction, and lid-driven cavity flow. The advantages of this high order gas-kinetic scheme are exhibited in some benchmarks including incompressible flow and supersonic compressible flow, inviscid flow and viscous flow.
\end{abstract}
	
	
\maketitle
	
\clearpage

\section{Introduction}
In the field of computational fluid dynamics (CFD), to design a robust, accurate numerical method for solving the hyperbolic conservation laws is a very attractive and active hot spots. In the scientific and engineering research area, many applications require the more accurate and resolved numerical approach \cite{liao2015extending}. The simulation of turbulent flows is a typical problem, and no matter the direct numerical simulation (DNS) or the large eddy simulation (LES) are all depended on the accuracy of the numerical scheme. Therefore, many kinds of high order schemes are developed, such as k-exact method \cite{barth1990higher}, essentially non-oscillatory (ENO) method \cite{abgrall1994essentially,durlofsky1992triangle,ollivier1997quasi,sonar1997construction}, weighted ENO (WENO) method \cite{liu1994weighted,shu1998essentially,hu1999weighted}, discontinuous Galerkin (DG) method \cite{cockburn2001runge}, and radial basis function method \cite{LIU20161096}. An excellent review of the high order methods can be referred to the work of Z. Wang \cite{wang2007high}.

Based on the idea of Bhatnagar et al. \cite{bhatnagar1954model}, the gas-kinetic scheme \cite{xu2001gas,xu2005multidimensional} is in the nature of spatial and temporal accuracy. It has been proved that gas-kinetic scheme is a robust and low dissipative numerical method. The advantages of gas-kinetic scheme have been recognized in the simulation of turbulent flows \cite{xiong2011numerical,FLD:FLD4239,li2016gas}, shock-boundary interaction and hypersonic flows \cite{li2005application,xu2005multidimensional}. A series studies based on the gas-kinetic scheme have been advanced, such as immersed boundary method \cite{yuan2015immersed,YUAN2018417}, implicit temporal marching \cite{li2014implicit}, and dual-time strategy \cite{PhysRevE.95.053307} for unsteady flows. But, for the high order method, gas-kinetic scheme is one of the fresh troops. The development of high order gas-kinetic scheme can be traced back to the study of Q. Li \cite{li2010high}. In the reference \cite{li2010high}, a novel high order scheme different from traditional method is developed. Following the idea of Q. Li, the high order gas-kinetic scheme should include two aspects of studies. One is the spatial reconstruction method, and the other is the high order gas-kinetic flux solver. In the field of gas-kinetic flux solver, Q. Li \cite{li2010high}, J. Luo \cite{luo2013high} and G. Zhou \cite{zhou2017simplification} make a foundational contribution, and an impressive two stage fourth order strategy is proposed by L. Pan \cite{pan2016efficient}. In the field of spatial reconstruction method, many algorithms are developed in the references \cite{luo2013high,pan2015third,pan2015compact} under various considerations, and the study of L. Pan \cite{pan2016third} is a compact method on unstructured grid. The impressive novelty in the work of \cite{pan2016third} is the conservative variables at the cell interface participating in the stage of spatial reconstruction. In present work, we focus on the spatial reconstruction method on unstructured grid, and for the characteristic of the gas-kinetic flux solver, present method can approach a third order accuracy without using multistage schemes.

For the high order finite volume methods, the accuracy and smoothness of the reconstruction polynomial are depended on the stencil of current cell seriously. The number of neighboring cells belonged to the stencil of current cell grows rapidly with the increase of the accuracy order. On the other hand, the parallel high performance computation requires the compactness of the reconstruction method. Thus, the high order finite volume methods always fall into the alternative decision of the accuracy or less stencil. CLS reconstruction method \cite{WANG2016863,WANG2016883} is designed to achieve arbitrary order accuracy under the finite volume method frame. The compactness of CLS method is of great promise in the high order method especially on unstructured grid. Since CLS method is compact and used all information from the whole computation domain implicitly, a large sparse linear system must be solved. To solve the linear system, the generalized minimal residual algorithm (GMRES) \cite{saad1986gmres,saad2003iterative} is applied in present work. The Reverse-Cuthill-McKee algorithm (RCM) \cite{cuthill1969reducing} and the incomplete lower-upper (ILU) preconditioning method \cite{meijerink1981guidelines,hackbusch1994iterative} are implemented to improve the efficiency of the solving procedure. In generally speaking, to solve the linear system occupies not a few computational costs on unstructured grid. But, compared with the expensive gas-kinetic flux solver, the procedure of solving linear system is not unacceptable.

The present paper is organized as follows. In Section~\ref{gks-flux}, the gas-kinetic flux solver is introduced briefly. The basic idea of CLS and construction of the linear system are introduced in Section~\ref{clsmethod}, and the distance weight biased averaging procedure (DWBAP) \cite{liu2017accuracy} is recommend. Several numerical cases are set up in the Section~\ref{cases}. The accuracy is proved to achieve the designed order, and the advantages of high order gas-kinetic scheme are exhibited in the cases. The latest section is a short conclusion.
\section{BGK equation and third order gas-kinetic flux solver}\label{gks-flux}
\subsection{BGK equation}
Based on the idea of Bhatnagar et al. \cite{bhatnagar1954model}, the Boltzmann equation can be expressed as
\begin{equation}\label{BGK}
f_t + \bm{u} \cdot \nabla{f} = -\frac{f-g}{\tau},
\end{equation}
where $\bm{u}$ is the particle velocity, $f$ is the distribution function of particles, $f_t$ represents the time derivative of $f$,
$\nabla f$ denotes the gradient of $f$, and $g$ is the equilibrium state
of Maxwellian distribution,
\begin{equation}\label{maxwellian}
g=\rho \left( \frac{\lambda}{\pi}\right)^{\frac{K+D}{2}}e^{-\lambda((\bm{u}-\bm{U}) \cdot (\bm{u}-\bm{U}) + \bm{\xi} \cdot \bm{\xi})},
\end{equation}
$D$ is the dimension, $\rho$ is the density, $\bm{U}$ represents the macroscopic fluid velocity,
$K$ is the total number of degrees of freedom in $\bm{\xi}$, $\bm{\xi}$ denotes the internal variables. Moreover, $\tau$ is the average
collision time, and $\lambda=m/2k_BT$, where $m$ is the molecular mass of particles, $k_B$ is the Boltzmann constant and $T$ is the
temperature. In this study $\lambda$ can be found from
\begin{equation} \label{lambda}
\lambda = \frac{(K+D)\rho}{4(E-0.5\rho \bm{U} \cdot \bm{U})},
\end{equation}
where $E$ denotes the energy of gas in the finite volume.

According to the gas dynamics theories, the macroscopic conservative variable $\bm{w}$ can be obtained by taking the moments of distribution function $f$ as follows
\begin{equation}\label{macroVar}
\bm{w} = \left(
	\begin{array}{c}
		\rho \\
		\rho \bm{u} \\
		E
	\end{array}
\right) = \int {\bm{\psi}fd\Xi},\quad \bm{\psi} = \left(1, \bm{u}, \frac{1}{2}\left(\bm{u} \cdot \bm{u} + \bm{\xi} \cdot \bm{\xi} \right) \right)^T,
\end{equation}
and the fluxes $\bm{F}$ at the cell interface can be read as
\begin{equation}\label{fluxes}
	\bm{F} = \left(
		\begin{array}{c}
			F_\rho \\
			F_{\rho \bm{u}} \\
			F_E
		\end{array}
\right) = \int {(\bm{u} \cdot \bm{n})\bm{\psi}fd\Xi},
\end{equation}
where $d\Xi = \left(\prod\limits_{i=1}^{D}du_i\right) \left(\prod\limits_{i=1}^{K}d\xi_i \right)$.
For a gas-kinetic scheme in finite volume method, the time dependent macroscopic conservative variable at the time
step $t_{n+1}$ can be expressed as
\begin{equation}\label{fvm3}
	\bm{w}_i^{n+1}=\bm{w}_i^{n} - \frac{1}{\|\Omega_i\|} \int_{t_n}^{t_{n+1}} \sum_{j=1}^{J}\bm{F}(t)Sdt,	
\end{equation}
where $i$  denotes the index of cells, $j$ means the index of interface belonged to the cell $i$, $J$ is the total number of the cell interfaces around, $\|\Omega\|$ is the measure of the control volume. In the high order finite volume method, the flux $\bm{F}$ is evaluated using the gauss integral at the cell interface to achieve the accuracy in space. For the cell interface which is perpendicular to the x-axis at $x_{i+1/2}$, the time dependent flux reads
\begin{equation}\label{fluxgauss}
	\bm{F}=\frac{1}{S}\sum_{l=1}^{N}\omega_l\bm{F}(x_{i+1/2}, y_l),
\end{equation}
where $S$ represents the measure of the cell interface and $w_l$ is related to the weight of gauss integral, and $N$ in Eq.~\ref{fluxgauss} represents the number of Gauss points.
\subsection{Third order gas-kinetic flux solver}
In this section, a third order gas-kinetic flux solver is introduced briefly.
For a time interval $[0, t]$, the integral form of general solution based on the characteristics at interface is given by Kogan \cite{GeneralSolution-Kogan}
\begin{equation}\label{generalSolution}
f(0,\bm{u}, t, \xi) = \frac{1}{\tau}\int_{0}^{t}g(\bm{x}', \bm{u}, t, \bm{\xi})e^{-(t-t')/\tau}dt'
+ e^{-t/\tau}f_0(-\bm{u}t),
\end{equation}
where $\bm{x}'=-\bm{u}(t-t')$, $f_0$ is the distribution function of particles at $t_0$.
The equilibrium state $g(0, \bm{u}, t)$ and initial non-equilibrium state
$f_0(0, \bm{u}, t)$ are connected together in Eq.~\ref{generalSolution}, and it implies that a distribution function at the cell interface can be decomposed as two parts, an equilibrium state and a non-equilibrium state.

For the BGK-NS scheme \cite{xu2001gas} in a 2D problem, the initial non-equilibrium state $f_{NS}$ around cell interface has the form
\begin{equation}\label{fNS}
	f_{NS}=g_k-\tau(ua_{1,k}+va_{2,k}+A_k)g_k, k=l,r,
\end{equation}
where, $a_{1,k}$, $a_{2,k}$ and $A_k$ denotes the normal, tangent and time derivative of initial distribution function from both sides of the cell interface respectively.  $u$ and $v$ are the particle velocities. $g_k$ is the Maxwellian distribution at the left or right of the cell interface. To construct a third order flux solver in two-dimensional, Eq.~\ref{fNS} should be expanded as
\begin{equation}\label{3orderf0:1}
	f_{0,k}=f_{NS}+\frac{\partial f_{NS}}{\partial x}x+\frac{\partial f_{NS}}{\partial y}y
	+\frac{1}{2}\frac{\partial^2 f_{NS}}{\partial x^2}x^2
	+\frac{\partial^2 f_{NS}}{\partial x\partial y}xy
	+\frac{1}{2}\frac{\partial^2 f_{NS}}{\partial y^2}y^2.
\end{equation}
Substituting Eq.~\ref{fNS} into Eq.~\ref{3orderf0:1} and ignoring the higher order derivatives (higher than two order), we can obtain the following formula
\begin{equation}\label{3orderf0:2}
	\begin{aligned}
		e^{-t/\tau}&f(-ut, y-vt, u, v) = c_7g_k\{1-\tau\left(a_{1,k}u+a_{2,k}v+A_k\right)\}\\
		&+c_8g_k\{
		a_{1,k}u-\tau\left(
		\left(a_{1,k}^2+a_{11,k}\right)u^2
		+\left(a_{1,k}a_{2,k}+a_{12,k}\right)uv
		+\left(A_ka_{1,k}+a_{1t}\right)u
		\right)
		\}\\
		&+c_8g_k\{
		a_{2,k}v-\tau\left(
		\left(a_{1,k}a_{2,k}+a_{12,k}\right)uv
		+\left(a_{2,k}^2+a_{22,k}\right)v^2
		+\left(A_ka_{2,k}+a_{2t}\right)v
		\right)
		\}\\
		&+c_7g_k\{
		a_{2,k}-\tau\left(
		\left(a_{1,k}a_{2,k}+a_{12,k}\right)u
		+\left(a_{2,k}^2+a_{22,k}\right)v
		+\left(A_ka_{2,k}+a_{2t}\right)
		\right)
		\}y\\
		&+\frac{1}{2}c_7g_k\{
		\left(a_{1,k}^2+a_{11,k}\right)\left(-ut\right)^2
		+2\left(a_{1,k}a_{2,k}+a_{12,k}\right)\left(-ut\right)\left(y-vt\right)
		+\left(a_{2,k}^2+a_{22,k}\right)\left(y-vt\right)^2
		\},
	\end{aligned}
\end{equation}
where
\begin{equation}\label{Nomenclature1}
	\begin{aligned}
	a_1 &= \frac{\partial g}{\partial x}/g,\quad a_2 = \frac{\partial g}{\partial y}/g, \quad
	A=\frac{\partial g}{\partial t}/g,\\
	a_{11} &= \frac{\partial a_1}{\partial x}, \quad 
	a_{12} = a_{21} = \frac{\partial a_1}{\partial y} = \frac{\partial a_2}{\partial x}, \quad 
	a_{22} = \frac{\partial a_2}{\partial y},\\
	a_{1t} &= \frac{\partial a_1}{\partial t} = \frac{\partial A}{\partial x}=A_1, \quad 
	a_{2t} = \frac{\partial a_2}{\partial t} = \frac{\partial A}{\partial y}=A_2, \quad 
	A_t=\frac{\partial A}{\partial t},\\
	\end{aligned}
\end{equation}
and
\begin{equation}\label{Nomenclature2}
\begin{aligned}
	c_1 &= 1-e^{-\frac{t}{\tau}},\quad	c_2=\left(t+\tau\right)e^{-\frac{t}{\tau}}-\tau,\quad c_3=t-\tau +\tau e^{-\frac{t}{\tau}},\\
	c_4 &= -e^{-\frac{t}{\tau}}\left(t^2+2t\tau\right),\quad c_5=t^2-2\tau t, \quad c_6=-t\tau-e^{-\frac{t}{\tau}}t\tau,\\
	c_7 &= e^{-t/\tau}, \quad c_8=-te^{-t/\tau}.
\end{aligned}
\end{equation}
To achieve the spatial and temporal accuracy of the scheme, the equilibrium state $g$ around the cell interface is assumed to have the form
\begin{equation}\label{3orderg:1}
	g=g_0+\frac{\partial g_0}{\partial x}x+\frac{\partial g_0}{\partial y}y+\frac{\partial g_0}{\partial t}t
	+\frac{1}{2}\frac{\partial^2 g_0}{\partial x^2}x^2+\frac{\partial^2 g_0}{\partial x\partial y}xy
	+\frac{1}{2}\frac{\partial^2 g_0}{\partial y^2}y^2+\frac{1}{2}\frac{\partial^2 g_0}{\partial t^2}t^2
	+\frac{\partial^2 g_0}{\partial x\partial t}xt+\frac{\partial^2 g_0}{\partial y\partial t}yt,
\end{equation}
where $g_0$ denotes the equilibrium state at the cell interface.
Combining with the Eq.~\ref{Nomenclature1}, Eq.~\ref{Nomenclature2} and Eq.~\ref{3orderg:1}, we can obtain the formula as follow
\begin{equation}\label{3orderg:2}
	\begin{aligned}
	\frac{1}{\tau}\int_{0}^{t}g(x', y', t', u, v)e^{-\frac{t-t'}{\tau}}dt' = & c_1g_0+c_2g_0\overline{a}_1u+c_2g_0\overline{a}_2v
	+c_1g_0\overline{a}_2y+c_3g_0\overline{A}\\
	&+\frac{1}{2}c_4g_0\left(\overline{a}_1^2+\overline{a}_{11}\right)u^2
	+c_6g_0\left(\overline{A}\overline{a}_1+\overline{A}_1\right)u
	+\frac{1}{2}c_5g_0\left(\overline{A}^2+\overline{A}_t\right)\\
	&+\frac{1}{2}c_1g_0\left(\overline{a}^2_2+\overline{a}_{22}\right)y^2
	+c_2g_0\left(\overline{a}^2_2+\overline{a}_{22}\right)vy
	+\frac{1}{2}c_4g_0\left(\overline{a}^2_2+\overline{a}_{22}\right)v^2 \\
	&+c_2g_0\left(\overline{a}_1\overline{a}_2+\overline{a}_{12}\right)uy
	+c_4g_0\left(\overline{a}_1\overline{a}_2+\overline{a}_{12}\right)uv \\
	&+c_3g_0\left(\overline{A}\overline{a}_2+\overline{A}_2\right)y
	+c_6g_0\left(\overline{A}\overline{a}_2+\overline{A}_2\right)v.
	\end{aligned}
\end{equation}
The symbols with $\bar{}$ in Eq.~\ref{3orderg:2} are related to the equilibrium state at the cell interface. The determination of various partial derivatives, such as $a_{1,k}$, $\overline{a}_1$ et al., can refer to the works of Q. Li \cite{li2010high}, J. Luo \cite{luo2013high}, and L. Pan \cite{pan2016third}. Substituting Eq.~\ref{3orderf0:2} and Eq.~\ref{3orderg:2}
into Eq.~\ref{generalSolution}, the flux across the interface can be computed using the formula Eq.~\ref{fluxgauss}.

\section{Compact least-square reconstruction}\label{clsmethod}
\subsection{Compact least-square reconstruction}
\begin{figure}[!htp]
	\centering
	\subfigure[Stencil for a triangular cell]{
		\includegraphics[width=0.40 \textwidth]{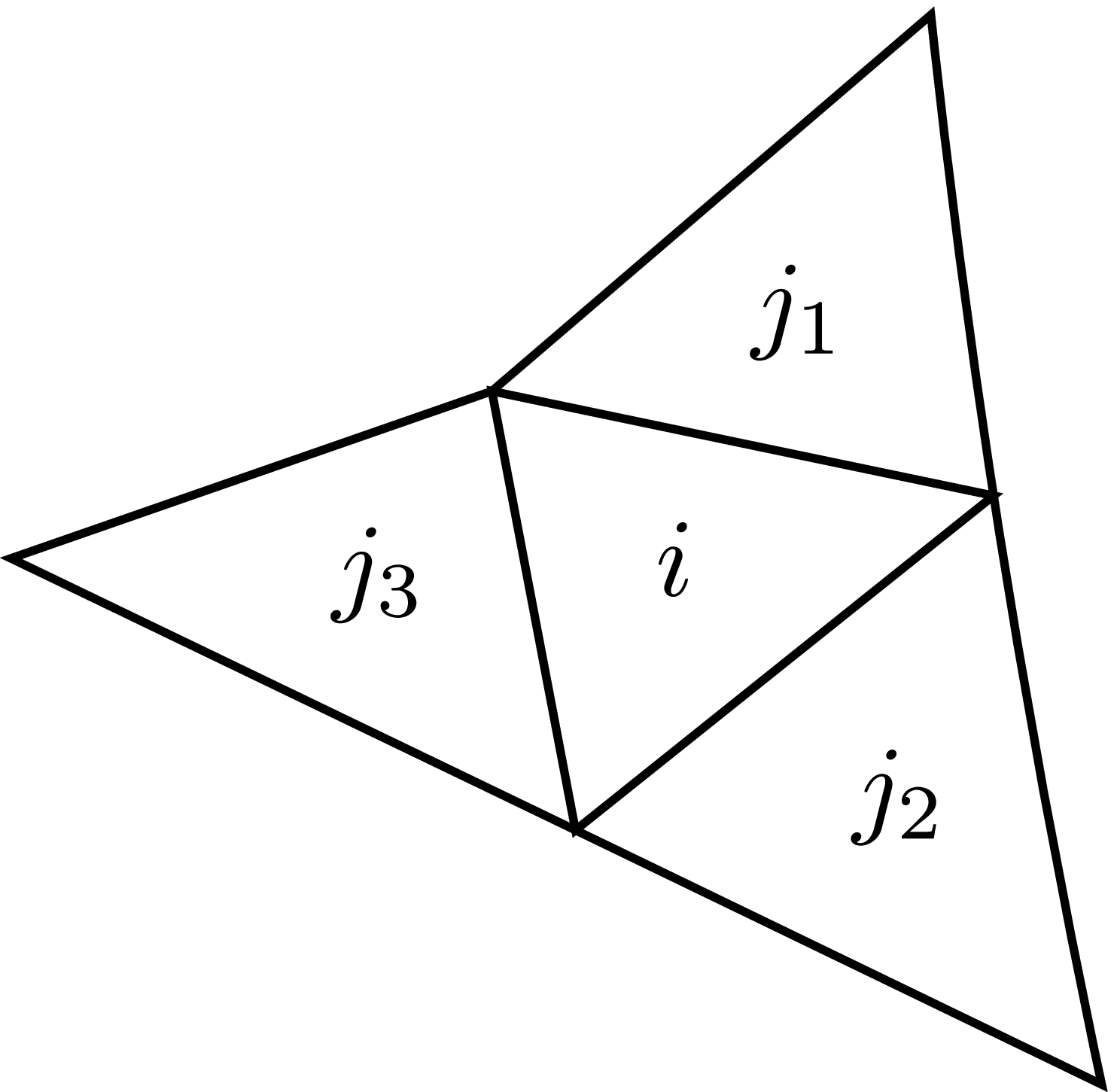}
		\label{fig:stencil:3}
	}
	\subfigure[Stencil for a quadrangular cell]{
		\includegraphics[width=0.40 \textwidth]{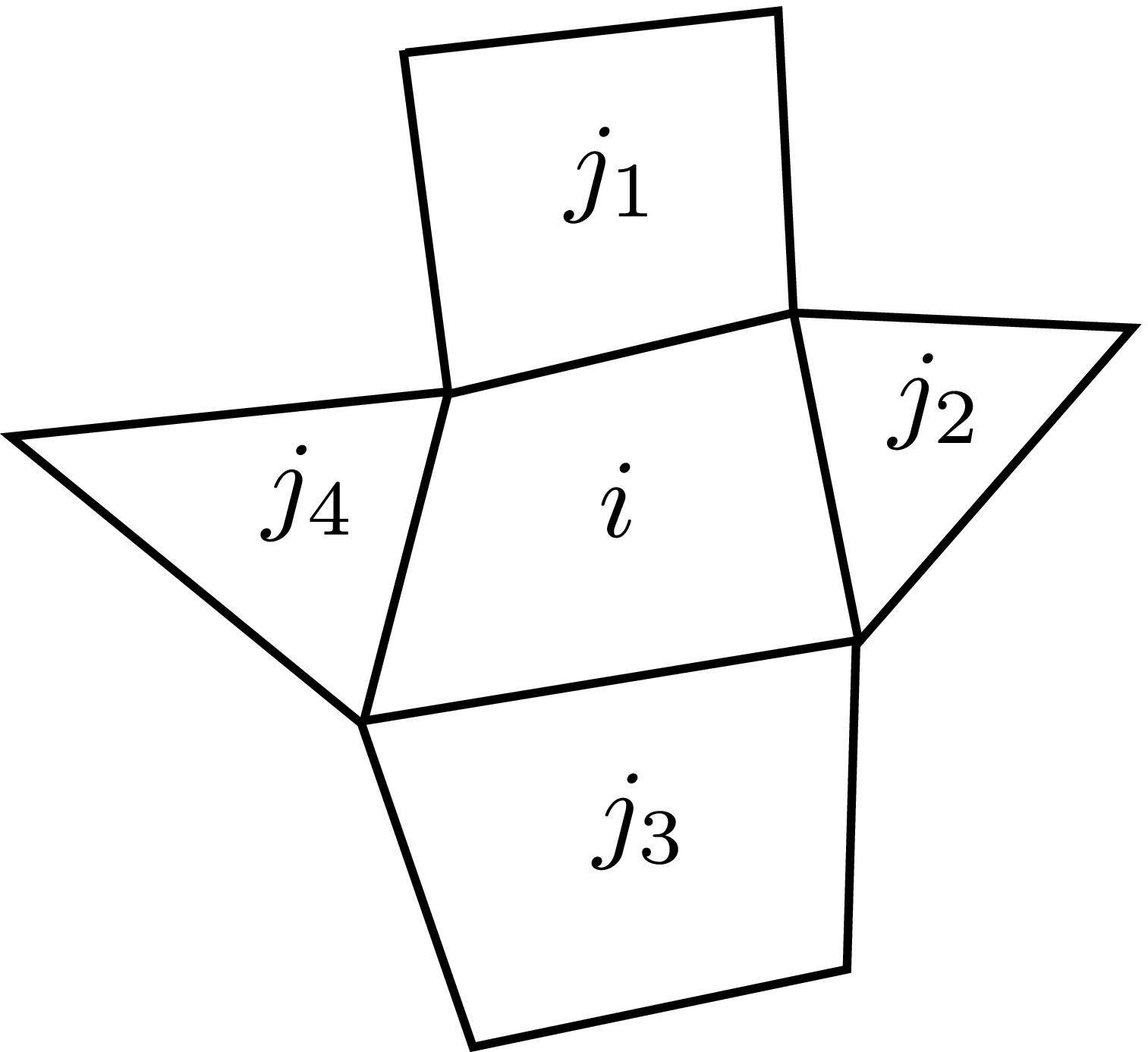}
		\label{fig:stencil:4}
	}
	\caption{Reconstruction stencil for cell i.}
	\label{fig:stencil}
\end{figure}

The CLS reconstruction method developed by Wang \cite{WANG2016863,WANG2016883} is based on the zero-mean basis, and it can be expressed as
\begin{equation}\label{eq:RCP}
	u^i(x,y)=\overline{u}^i+\sum_{l=1}^{DOF(k)}u^i_l\phi_{l,i}(x, y),
\end{equation}
where
\begin{equation*}
	\begin{aligned}
		\overline{u}^i &= \frac{1}{\|\Omega_i\|}\int_{\Omega_i}u(\bm{x})d\Omega,
		\quad \phi_{l,i}(x,y) = \Delta x^m_i\Delta y^n_i-\overline{\Delta x^m_i\Delta y^n_i},\\
		\Delta x_i &= \frac{x-x_i}{h_i},\quad \Delta y_i=\frac{y-y_i}{h_i}, \quad
		\overline{\Delta x^m_i\Delta y^n_i}=\frac{1}{\|\Omega_i\|}\int_{\Omega_i}\Delta x^m_i\Delta y^n_id\Omega.
	\end{aligned}
\end{equation*}
We emphasize that the symbol $u$ used in this section represents the variable which is to be reconstructed. $h_i$ denotes the length scale for the non-dimensionalization of the basis functions to avoid growth of the condition number of the reconstruction matrix with grid refinement. In present work, the length scale is defined as
\begin{equation}\label{eq:h}
	h_i=max(Rad_i, \sqrt{\|\Omega_i\|}),
\end{equation}
where $Rad_i$ is the radius of the circumcircle of the control volume $i$. The freedom of $k$ order polynomial reads
\begin{equation}\label{dof_k}
	DOF(k)=\left(k+1\right)\left(k+2\right)/2-1.
\end{equation}
According to Eq.~\ref{eq:RCP}, a quadratic reconstruction $(k=2)$ polynomial can be rewritten as
\begin{equation}\label{eq:quadRCP}
	\begin{aligned}
		u^i(x,y) = \overline{u}^i &+ \sum_{l=1}^{5}u^i_l\phi_{l,i}(x, y)=\overline{u}^i+u^i_1\Delta x+u^i_2\Delta y\\
		&+\frac{1}{2}u^i_3\left(\Delta x^2 - \overline{\Delta x^2_i}\right)
		+u^i_4\left(\Delta x\Delta y - \overline{\Delta x_i\Delta y_i}\right)
		+\frac{1}{2}u^i_5\left(\Delta y^2 - \overline{\Delta y^2_i}\right).
	\end{aligned}
\end{equation}
Because of the use of zero-mean basis, the following condition,
\begin{equation}\label{eq:cellAVGCond}
	\overline{u}^i=\frac{1}{\|\Omega_i\|}\int_{\Omega_i}u(x,y)d\Omega,
\end{equation}
is always satisfied in nature. However, to determine the free parameters $u^i_l$ in Eq.~\ref{eq:quadRCP}, more other equations related to the stencil for the cell $i$ must be added. Fig.~\ref{fig:stencil} shows the stencil used in the spatial reconstruction on control volume $i$.

In the CLS method, the various orders of spatial derivatives of reconstruction polynomial $u(x,y)$ are required to be conserved on $S_i$ ($S_i=\{\Omega_1, \Omega_2, \cdots, \Omega_J\}$, $J$ denotes the total number of control volumes neighboring the cell $i$). Namely, for all $\Omega_j \in S_i$,
\begin{equation}\label{eq:spdConserved:1}
	\frac{1}{\|\Omega_j\|}\int_{\Omega_j}\frac{\partial^{m+n}u^i(x,y)}{\partial x^my^n}dxdy=\frac{1}{\|\Omega_j\|}\int_{\Omega_j}\frac{\partial^{m+n}u^j(x,y)}{\partial x^my^n}dxdy, \quad 0\leq m+n \leq M,
\end{equation}
where $M \leq k$. Substituting Eq.~\ref{eq:RCP} into Eq.~\ref{eq:spdConserved:1}, we obtain the following linear equations
\begin{equation}\label{eq:spdConserved:2}
	\begin{aligned}
		\sum_{l=1}^{DOF(k)}u^i_l\left(\frac{1}{\|\Omega_j\|}\int_{\Omega_j}\frac{\partial^{m+n}\phi_{l,i}(x,y)}{\partial x^my^n}dxdy\right)
		= & \delta^0_{m+n}\left(\overline{u}^j-\overline{u}^i\right)\\
		& +\sum_{l=1}^{DOF(k)}u^j_l\left(\frac{1}{\|\Omega_j\|}\int_{\Omega_j}\frac{\partial^{m+n}\phi_{l,j}(x,y)}{\partial x^my^n}dxdy\right).
	\end{aligned}
\end{equation}
Let
\begin{equation}\label{eq:u_i}
	\bm{u}^i=\left(u^i_1, u^i_2, \cdots ,u^i_{DOF(k)}\right)^T,
\end{equation}
Eq.~\ref{eq:spdConserved:2} can be rewritten as
\begin{equation}\label{eq:spdConserved:3}
	\bm{A}^i_j\bm{u}^i-\bm{B}^i_j\bm{u}^j=\bm{b}^i_j,
\end{equation}
where
\begin{equation}\label{eq:A:B:b}
	\begin{aligned}
	\bm{A}^i_j &= \left[\frac{1}{\|\Omega_j\|}\int_{\Omega_j}\frac{\partial^{m+n}\phi_{l,i}(x,y)}
		{\partial x^my^n}dxdy\right]_{(DOF(M)+1) \times DOF(k)},\\
	\bm{B}^i_j &= \left[\frac{1}{\|\Omega_j\|}\int_{\Omega_j}\frac{\partial^{m+n}\phi_{l,j}(x,y)}
		{\partial x^my^n}dxdy\right]_{(DOF(M)+1) \times DOF(k)},\\
	\bm{b}^i_j &= \left[\delta^0_{m+n}\left(\overline{u}_j-\overline{u}_i\right)\right]_{(DOF(M)+1) \times 1}.
	\end{aligned}
\end{equation}

According to Gustafsson \cite{gustafsson1975convergence}, the designed accuracy can be achieved with the accuracy at the boundary keeps one order lower than the interior of the computational domain. Therefore, the Eq.~\ref{eq:RCP} and Eq. ~\ref{eq:A:B:b} can be rewritten as
\begin{equation}\label{eq:RCP:2}
u^i(x,y)=\overline{u}^i+\sum_{l=1}^{DOF(k_i)}u^i_l\phi_{l,i}(x, y),
\end{equation}
\begin{equation}\label{eq:A:B:b:2}
	\begin{aligned}
	\bm{A}^i_j &= \left[\frac{1}{\|\Omega_j\|}\int_{\Omega_j}\frac{\partial^{m+n}\phi_{l,i}(x,y)}
	{\partial x^my^n}dxdy\right]_{(DOF(M)+1) \times DOF(k_i)},\\
	\bm{B}^i_j &= \left[\frac{1}{\|\Omega_j\|}\int_{\Omega_j}\frac{\partial^{m+n}\phi_{l,j}(x,y)}
	{\partial x^my^n}dxdy\right]_{(DOF(M)+1) \times DOF(k_j)},\\
	\bm{b}^i_j &= \left[\delta^0_{m+n}\left(\overline{u}_j-\overline{u}_i\right)\right]_{(DOF(M)+1) \times 1},
	\end{aligned}
\end{equation}
where the $k_i$ is the order of the reconstruction polynomial. For a third order scheme,
\begin{equation}\label{k_i}
k_i = \left\{
\begin{array}{l}
	1, \quad \text{cell i is at the boundary},\\
    2, \quad \text{cell i is in the interior}.
\end{array}
\right.
\end{equation}
Following the advices of the reference \cite{WANG2016883}, the $\bm{A}^i_j$ and $\bm{B}^i_j$ are associated with a weight function $w_{i,p}$ to adjust the effect of different order of partial derivatives.
\begin{equation}\label{eq:A:B:b:3}
\begin{aligned}
\bm{A}^i_j &= \left[\frac{w_{i,p}}{\|\Omega_j\|}\int_{\Omega_j}\frac{\partial^{m+n}\phi_{l,i}(x,y)}
{\partial x^my^n}dxdy\right]_{(DOF(M)+1) \times DOF(k_i)},\\
\bm{B}^i_j &= \left[\frac{w_{i,p}}{\|\Omega_j\|}\int_{\Omega_j}\frac{\partial^{m+n}\phi_{l,j}(x,y)}
{\partial x^my^n}dxdy\right]_{(DOF(M)+1) \times DOF(k_j)},
\end{aligned}
\end{equation}
where, $w_{i,p}=(wh_i)^p$, $p=m+n$, and $w=0.3$ is chosen in our work. It must be emphasized that the weight function is of great importance for the linear system to be solved. An improper weight function could bring much difficulty to solve the linear system.
\subsection{The solving procedure of linear system}
CLS method is a compact reconstruction scheme used the stencil from the global computational domain implicitly. According to  Eq.~\ref{eq:spdConserved:3}, a large sparse linear system  constructed from unstructured grid must be solved. Let
\begin{equation}\label{eq:u}
	\bm{u}^i_j = \{\cdots, \bm{u}^i, \cdots, \bm{u}^j,\cdots\}^T,
\end{equation}
\begin{equation}\label{eq:cij}
\bm{C}^i_j = \{\cdots, \bm{A}^i_j, \cdots, \bm{B}^i_j, \cdots \},
\end{equation}
Eq.~\ref{eq:spdConserved:3} can be rewritten as
\begin{equation}\label{eq:spdConserved:4}
	\bm{C}^i_j\bm{u}^i_j=\bm{b}^i_j.
\end{equation}
In practice, the over-determined linear system described by Eq.~\ref{eq:spdConserved:4} is solved using the least-square method. The corresponding normal equations read
\begin{equation}\label{eq:spdConserved:5}
	\left(\bm{C}^i_j\right)^T\bm{C}^i_j\bm{u}^i_j=\left(\bm{C}^i_j\right)^T\bm{b}^i_j.
\end{equation}

 But, due to the unstructured grid, the nonzero elements in the sparse matrix are distributed uncontrollably and undesirably. It widens the band width of the sparse matrix and slows down the convergence of iterative method. In order to reduce band width of sparse matrix and relieve the harm of the random distribution of nonzero elements, the renumbering strategy, Reverse-Cuthill-Mckee algorithm \cite{cuthill1969reducing} (RCM), is used in our study.

Following the renumbering stage, the GMRES \cite{saad1986gmres,saad2003iterative} algorithm is applied to solve the linear system. In order to improve the efficiency of iterative method, the incomplete lower upper factorization method \cite{meijerink1981guidelines,hackbusch1994iterative} is implemented to cluster the eigenvalues of the system matrix.

\subsection{DWBAP high order limiter}
To suppress the non-physical oscillations near the discontinuities, a distance weighted biased averaging procedure (DWBAP) developed by Liu \cite{liu2017accuracy} is applied. DWBAP is proved to be an accuracy preserving limiter for high order finite volume method on unstructured grid, and it can be expressed as
\begin{equation}\label{DWBAP}
	L(u^0_l, u^1_l, \cdots, u^J_l) = B^{-1}\left(\sum_{j=0}^{J}\omega_jB(u^i_j)\right),
	\quad B(x)=arctan(x), \quad B^{-1}(x)=tan(x),
\end{equation}
where $J$ represents the total number of neighbors adjacent to the control volume $i$. $u^0_l$ is the variable at cell $i$ to be limited. $j \in [1, J]$ is related to the neighbors of cell $i$. $\omega$ denotes the weighting coefficient of biased function $B(x)$, which is defined as follows
\begin{equation}\label{DWBAP_w}
	\omega_j = \frac{\beta_j}{\sum_{j=0}^{J}\beta_j}, \quad \beta_j=\frac{1}{\left(\frac{1}{J}\sum_{m=1}^{J}|r_j-r_m|\right)^{S(\chi)}}.
\end{equation}
$r_j$ is the centroid of the neighboring cell, and $r_m$ is the face centroid of cell $i$. $S(\chi)$ is a smooth function, which is defined as
\begin{equation}\label{DWBAP_S}
	S(\chi) = \frac{1}{\sqrt{\chi+\epsilon}+0.05}, \quad \chi = \frac{\sigma}{\overline{X}},
	\quad \sigma = \sqrt{\frac{1}{J}\sum_{j=1}^{J}\left(X_j-\overline{X}\right)^2},
\end{equation}
where $\epsilon$ is chosen as $10^{-8}$. $X$ is a specific variable to evaluate the smoothness around cell $i$, and pressure $p$ is used in our paper.
To improve the efficiency of limiting procedure, a problem independent shock detector is introduced as follows
\begin{equation}\label{ISD1}
	IS_i = \frac{\sum_{j=1}^{J}|u_i(r_i)-u_j(r_i)|}{Jh^{(k+1)/2}_i max(\overline{u}_j, \overline{u}_i)},
\end{equation}
and
\begin{equation}\label{ISD2}
	\left\{
		\begin{array}{l}
			IS_i < 1, \textbf{smooth region}, \\
			IS_i \geq 1, \textbf{shock region}.
		\end{array}
	\right.
\end{equation}

\section{Numerical results}\label{cases}
In this section, numerical tests are set up for the validation of present method. The cases includes both inviscid and viscous flows, incompressible and compressible flows. For the inviscid flow, the collision time $\tau$ is
\begin{equation}\label{tauIns}
	\tau = 0.01\Delta t+\frac{|p_l-p_r|}{|p_l+p_r|}.
\end{equation}
For the viscous flow, the collision time $\tau$  has form
\begin{equation}\label{tauVis}
	\tau = \frac{\mu}{p}+\frac{|p_l-p_r|}{|p_l+p_r|},
\end{equation}
where $p_l$ and $p_r$ are the pressure computed from the state of both sides of the cell interface. $\mu$ represents the dynamic viscosity. $p$ denotes the pressure at the cell interface.

It should be noted that we use the length scale $h$ to describe the cell size of the grid used in the simulation. $h$ represents the measure of division at the bound of computational domain in this section.
\subsection{Accuracy tests}
\begin{figure}[!htp]
	\centering
	\subfigure[$L_2$ in density vs. $h$]{
		\includegraphics[width=0.45 \textwidth]{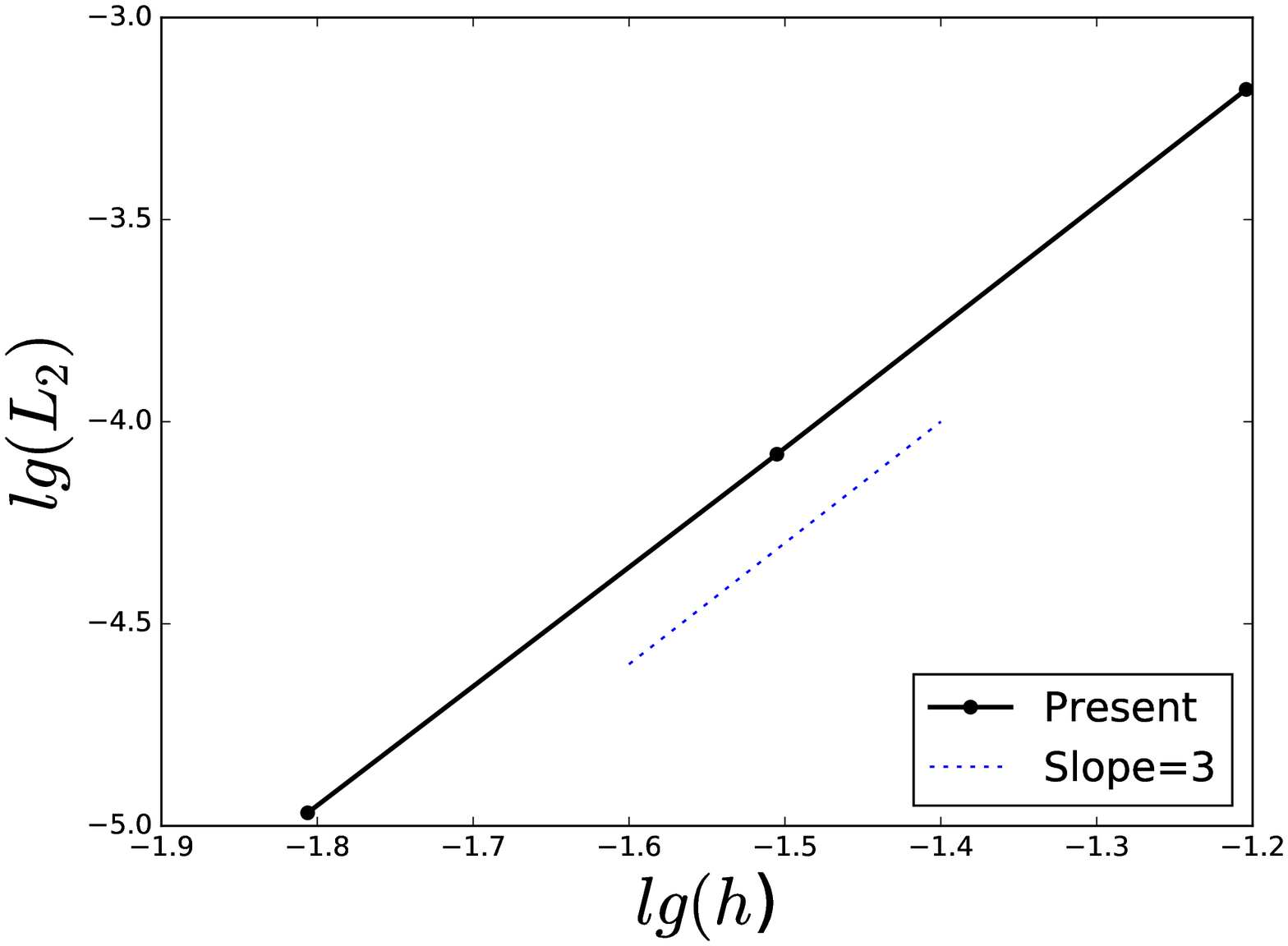}
		\label{fig:denAdvOrder}
	}
	\subfigure[Error distribution with different $h$]{
		\includegraphics[width=0.45 \textwidth]{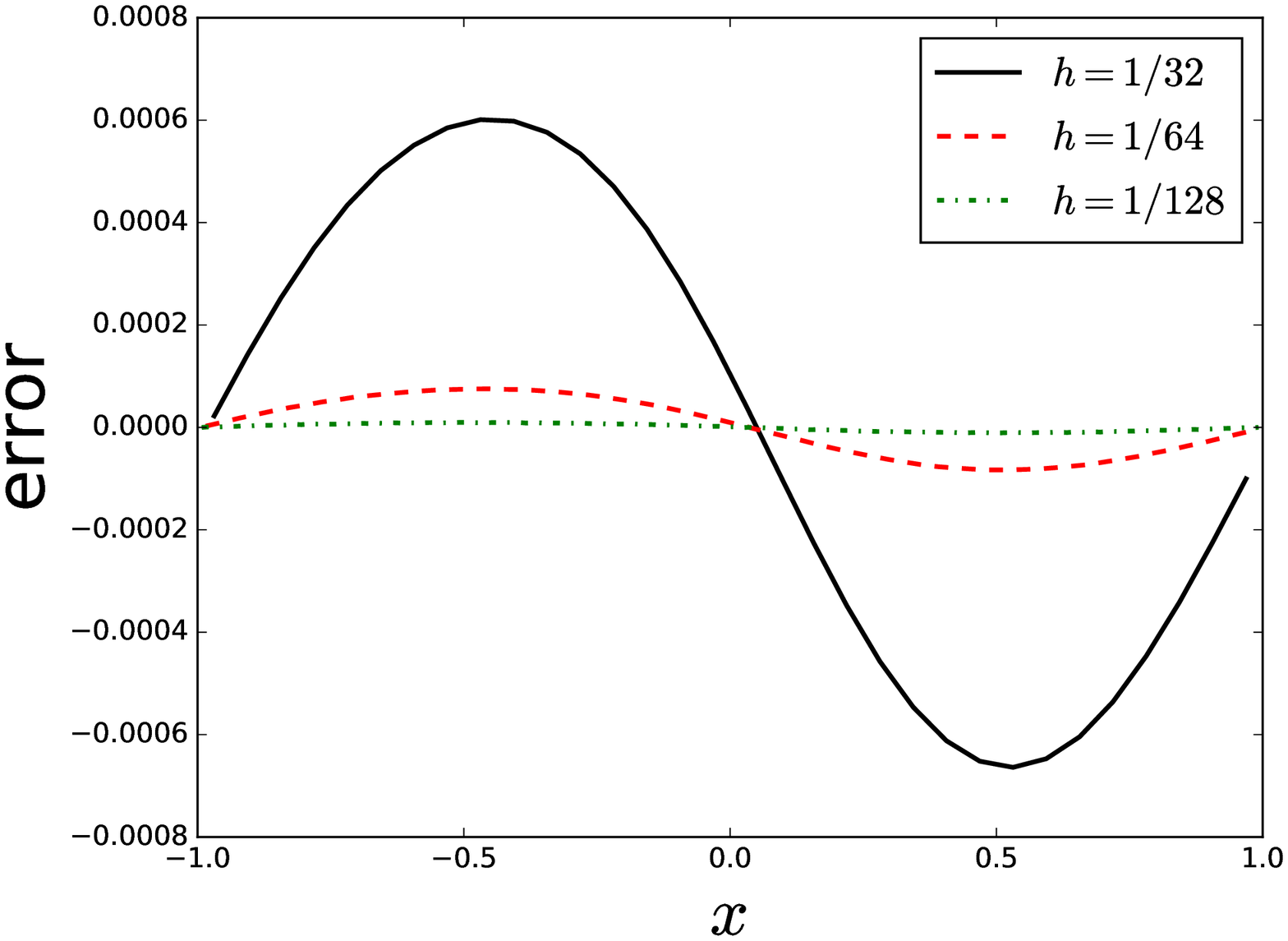}
		\label{fig:denAdvError}
	}
	\caption{The results of accuracy test.}
	\label{fig:denAdv}
\end{figure}

In this case, the advection of density perturbation problem is presented to validate the accuracy of our method.
The initial condition is given as
\begin{equation}\label{InitDenPerturbation}
	\rho(x) = 1+0.2sin(\pi x), \quad u(x) = 1, \quad v(x) = 0, \quad p(x) = 1,
\end{equation}
and the analytic solution at the time $t$ can be expressed as
\begin{equation}\label{SolDenPerturbation}
	\rho(x, t) = 1+0.2sin(\pi (x-t)), \quad u(x, t) = 1, \quad v(x, t) = 0, \quad p(x, t) = 1.
\end{equation}
The case is a one-dimensional problem, and we simulate it using a two-dimensional solver on a  Cartesian grid. The computational
domain is
\begin{equation}
	\{(x,y)| x \in [0,1], y \in [0, 4h]\}.
\end{equation}
The periodic boundary conditions is implemented at the corresponding bound of the computational domain. The numerical results are shown in Fig.\ref{fig:denAdv}. It can be concluded that the accuracy designed is achieved in our method, and the error distribution of density shows an excellent performance of the accuracy increase with the mesh resolution.

\subsection{Sod problem}
\begin{figure}[!htp]
	\centering
	\subfigure[Structured grid]{
		\includegraphics[width=0.9 \textwidth]{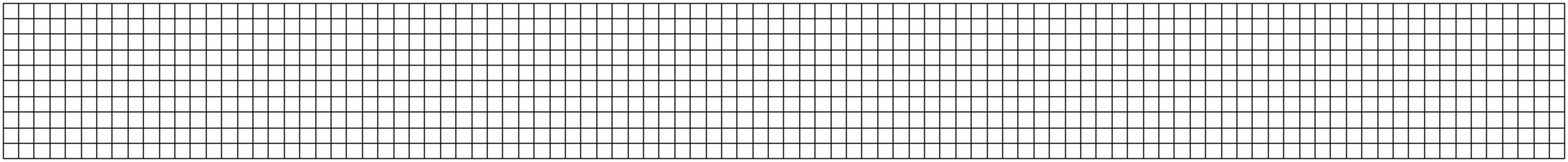}
		\label{fig:sod:structured}
	}
	\subfigure[Unstructured grid 1]{
		\includegraphics[width=0.9 \textwidth]{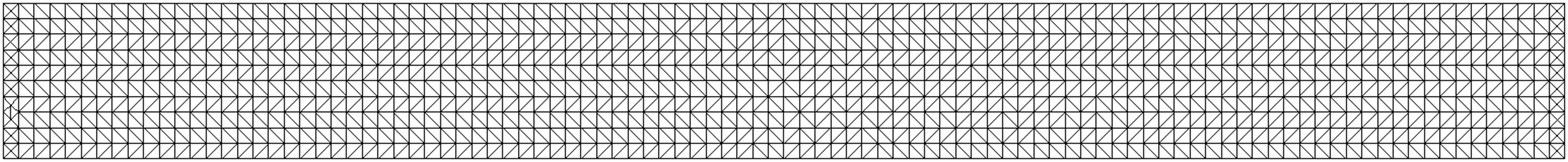}
		\label{fig:sod:unstructured1}
	}
	\subfigure[Unstructured grid 2]{
		\includegraphics[width=0.9 \textwidth]{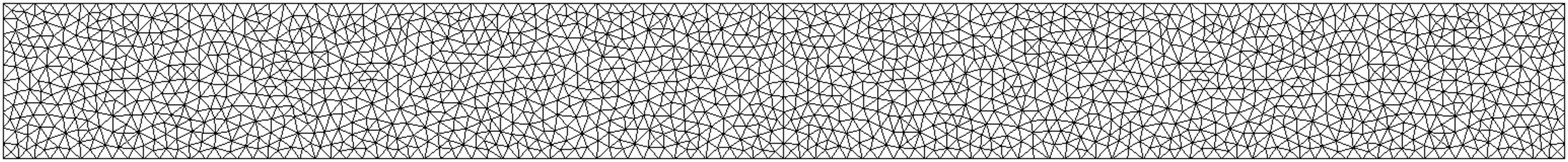}
		\label{fig:sod:unstructured2}
	}
	\caption{The grid used in the computation of Sod problem.}
	\label{fig:sodGrid}
\end{figure}

\begin{figure}[!htp]
	\centering
	\subfigure[$\rho$]{
		\includegraphics[width=0.3 \textwidth]{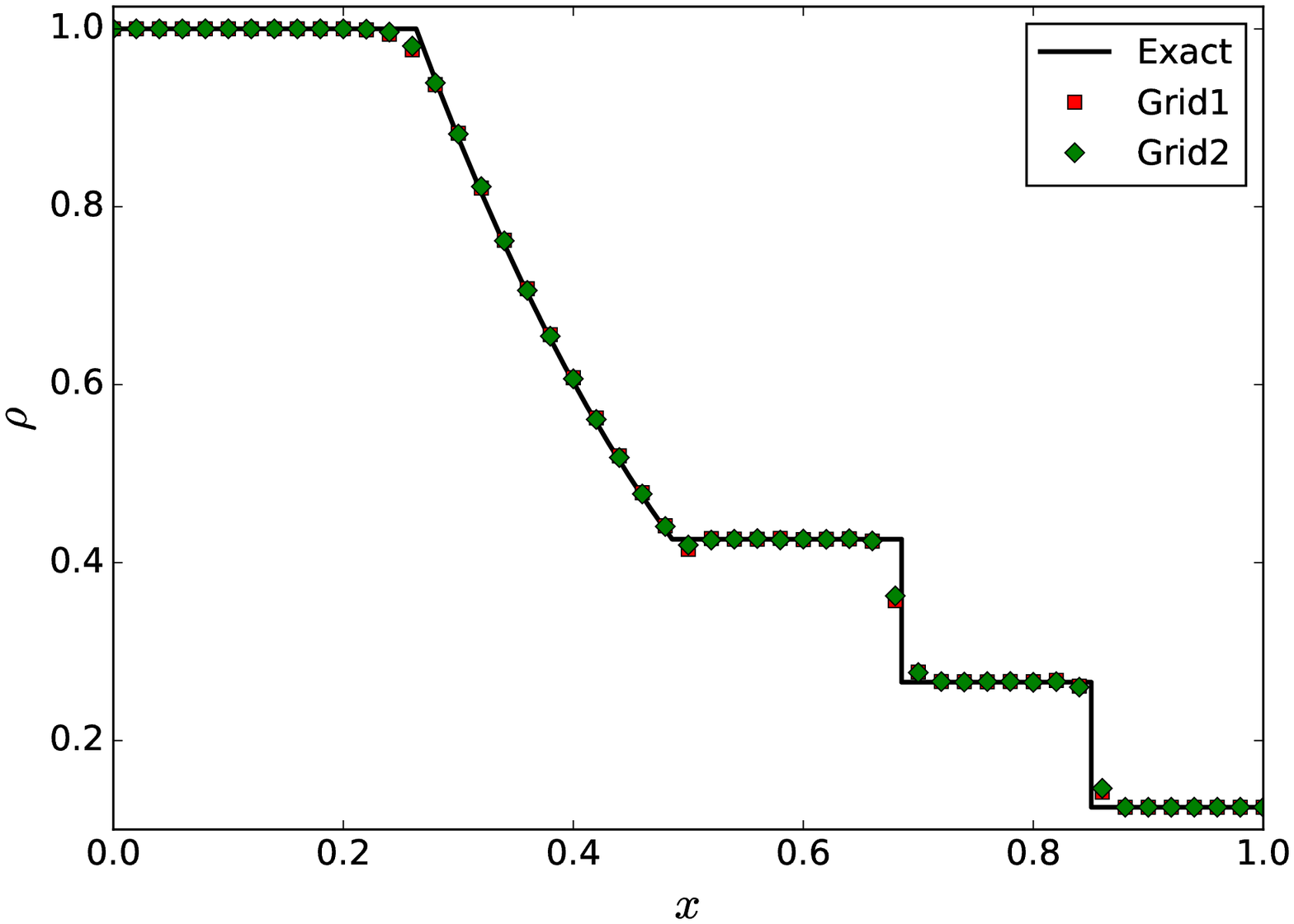}
		\label{fig:sod:denUnstructured}
	}
	\subfigure[$u$]{
		\includegraphics[width=0.3 \textwidth]{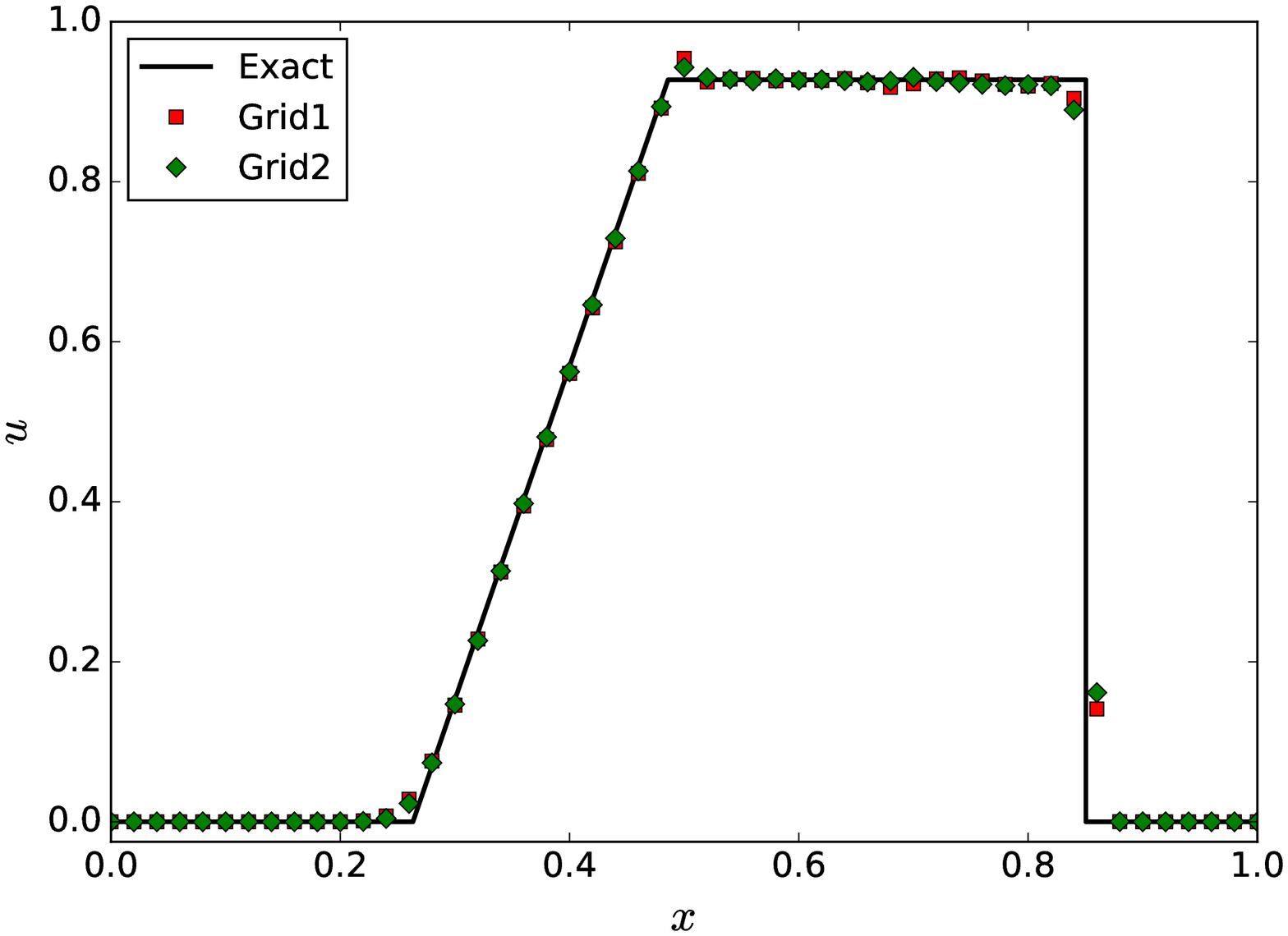}
		\label{fig:sod:uUnstructured}
	}
	\subfigure[$p$]{
		\includegraphics[width=0.3 \textwidth]{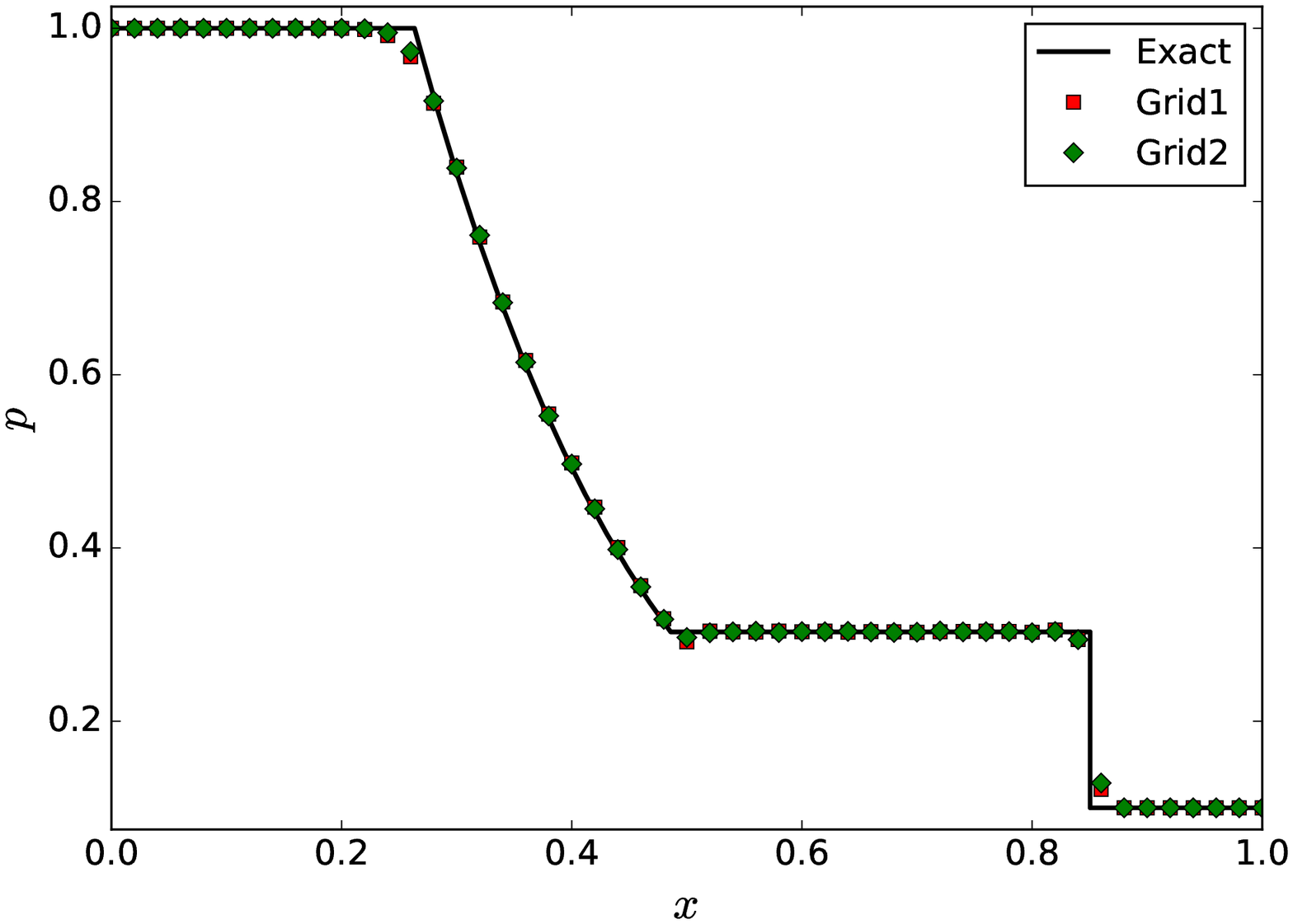}
		\label{fig:sod:pUnstructured}
	}
	\caption{The numerical results of Sod problem on unstructured grid.}
	\label{fig:sodUnstructuredGrid}
\end{figure}

Sod problem is a one-dimensional Riemann problem, which is always used to validate the ability of numerical schemes to capture the discontinuity. The initial condition reads
\begin{equation}\label{sodInit}
	\left(\rho, u, v, p\right)= \left\{
		\begin{array}{ll}
			\left(1, 0, 0, 1\right),& 0<x<0.5,\\
			\left(0.125, 0, 0, 0.1\right), & 0.5 \leq x \leq 1.
		\end{array}
		\right.
\end{equation}
In this case, both the Cartesian grid and unstructured grid are used, and the computational domain is $[0,1]\times[0,0.1]$. The grids used in the computation are shown in Fig.~\ref{fig:sodGrid}, and the length scale $h$ of the grids is $1/100$. The distribution of density, velocity and pressure are shown in Fig.~\ref{fig:sodUnstructuredGrid}. Both the results from two grids have good accordance with the exact solution.

\begin{figure}[!htp]
	\centering
	\subfigure[$\rho$]{
		\includegraphics[width=0.3 \textwidth]{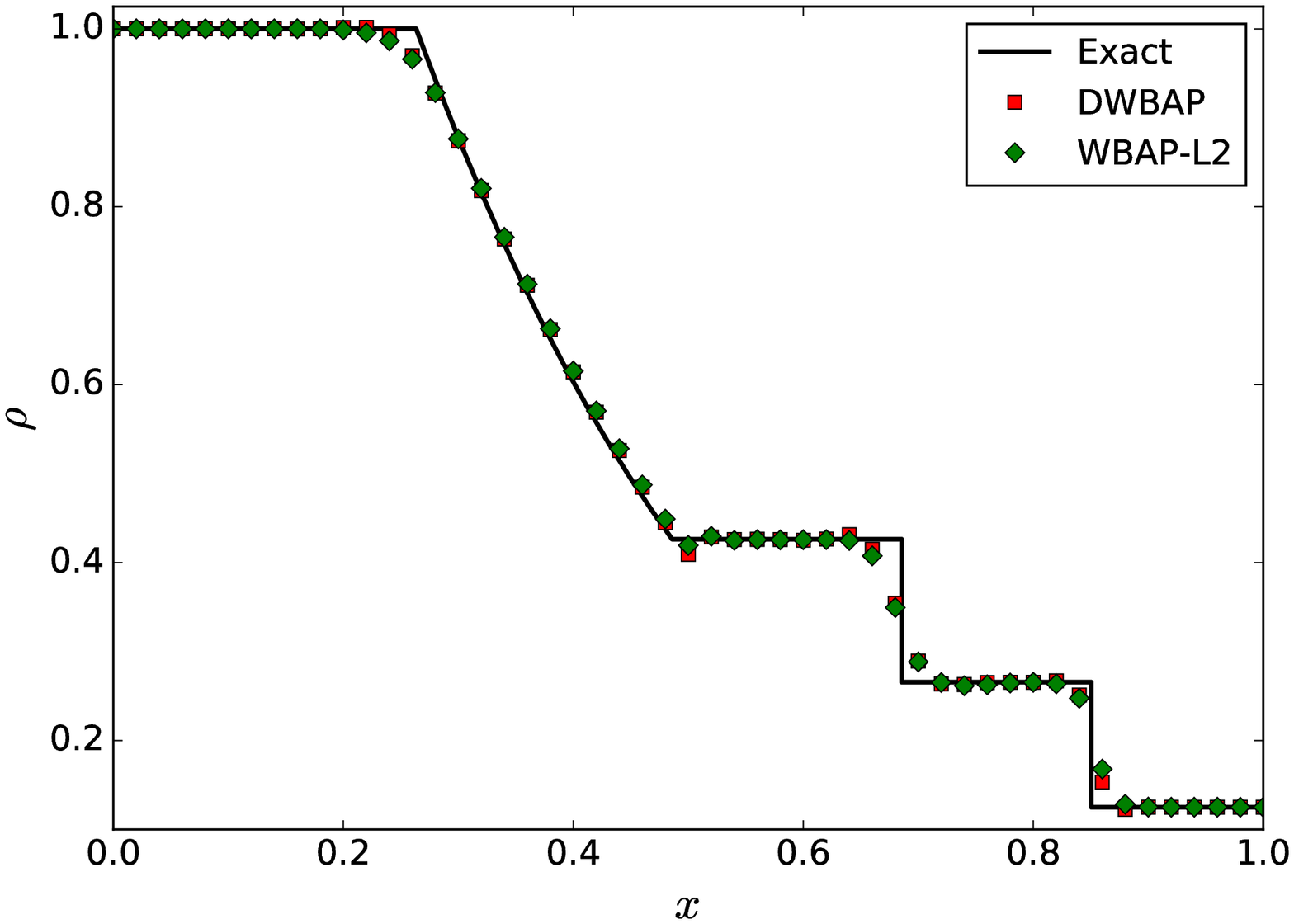}
		\label{fig:sod:denStructured}
	}
	\subfigure[$u$]{
		\includegraphics[width=0.3 \textwidth]{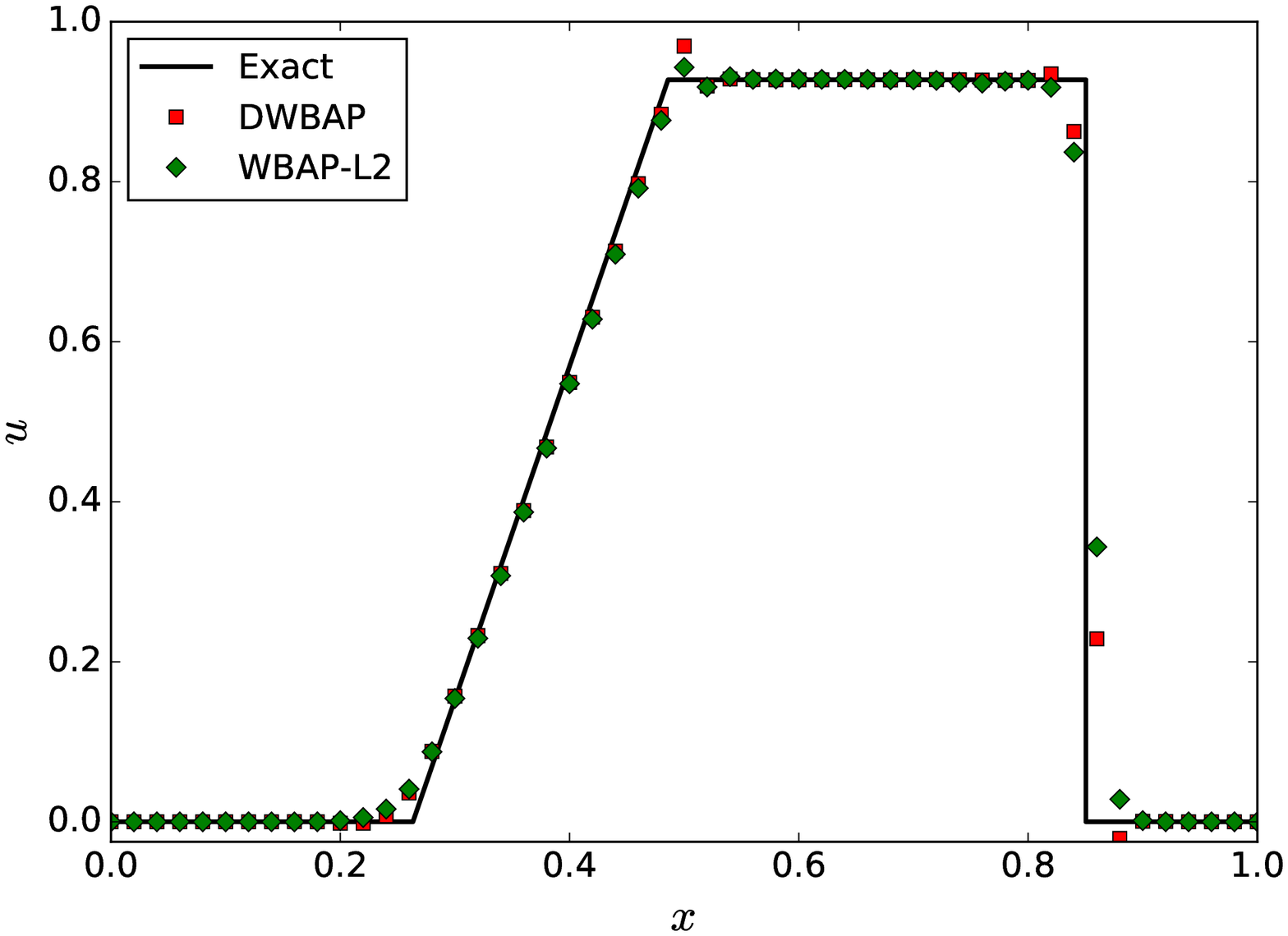}
		\label{fig:sod:uStructured}
	}
	\subfigure[$p$]{
		\includegraphics[width=0.3 \textwidth]{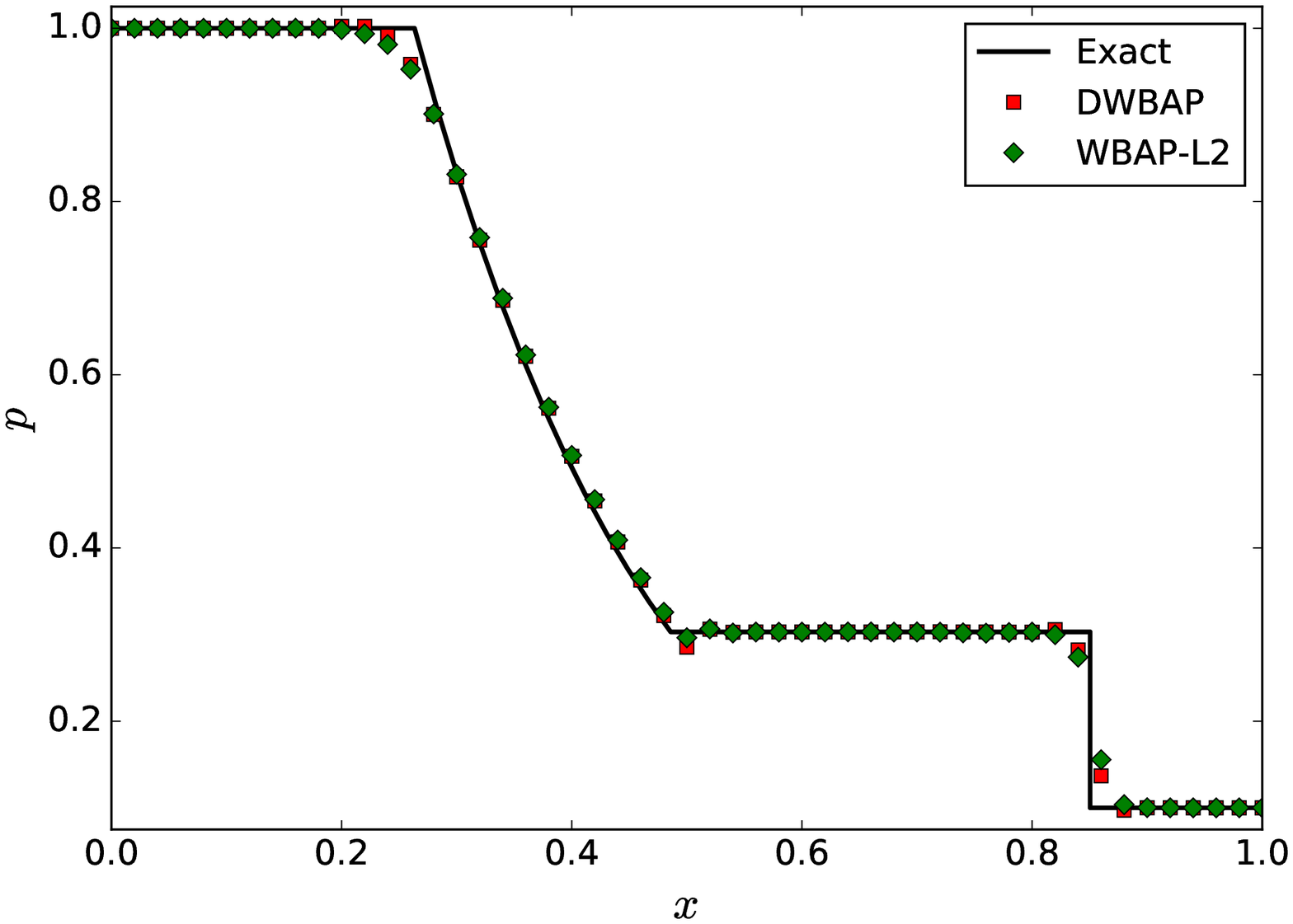}
		\label{fig:sod:pStructured}
	}
	\caption{The effects of limiters in terms of conservative variables on structured grid.}
	\label{fig:sodLimiter}
\end{figure}

\begin{figure}[!htp]
	\centering
	\subfigure[$\rho$]{
		\includegraphics[width=0.3 \textwidth]{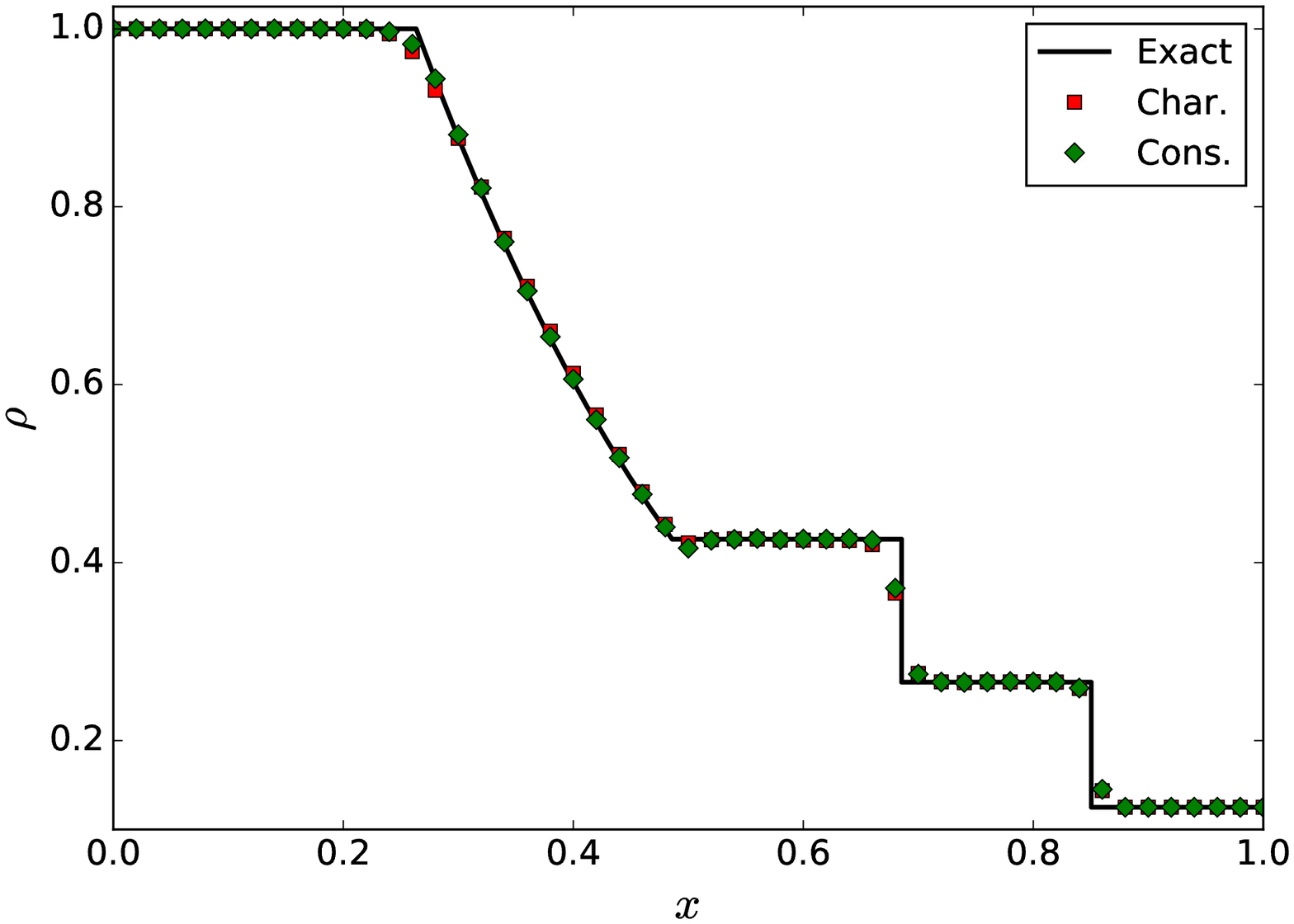}
		\label{fig:sod:denUnstructuredCharVsCons}
	}
	\subfigure[$u$]{
		\includegraphics[width=0.3 \textwidth]{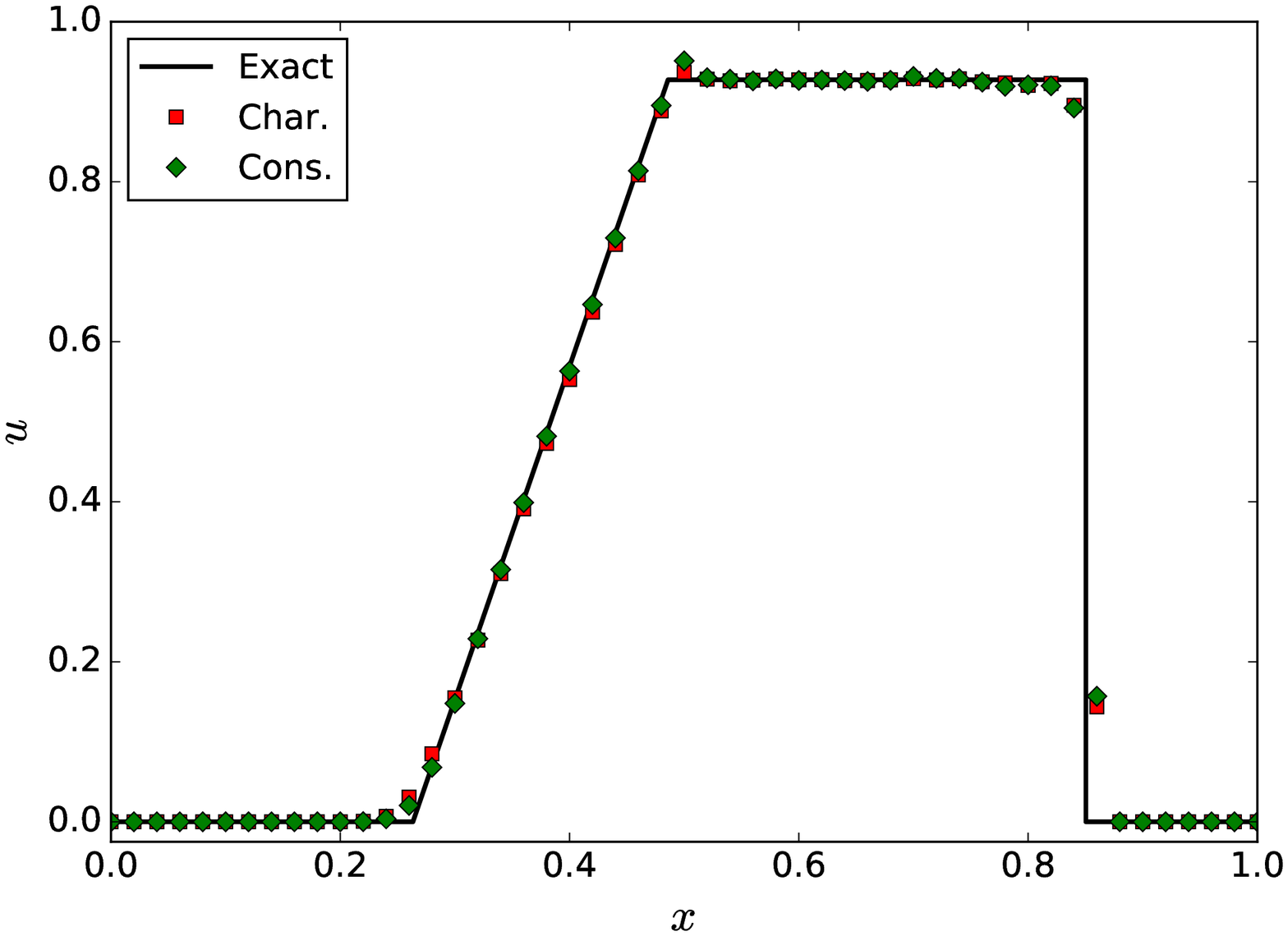}
		\label{fig:sod:uUnstructuredCharVsCons}
	}
	\subfigure[$p$]{
		\includegraphics[width=0.3 \textwidth]{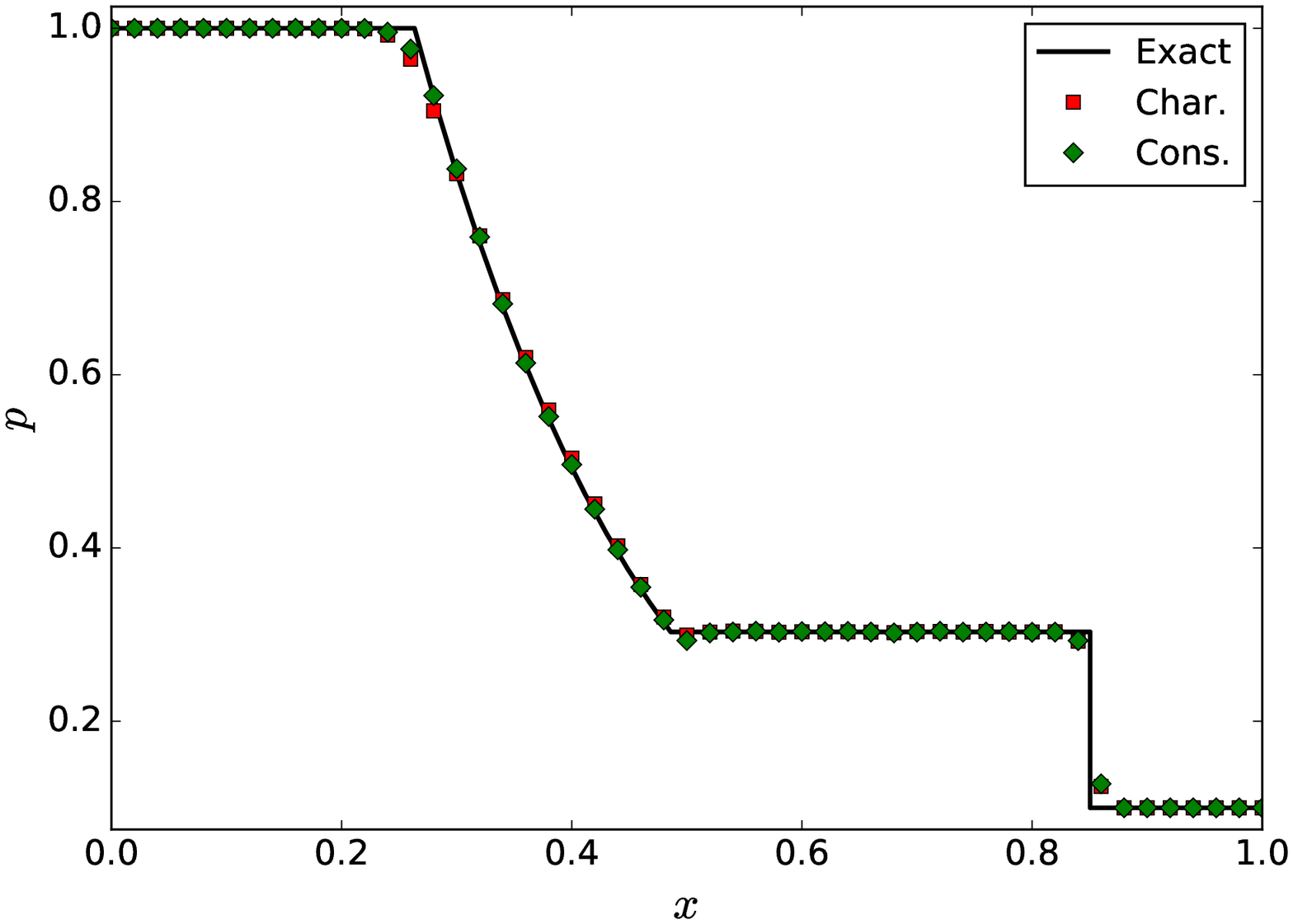}
		\label{fig:sod:pUnstructuredCharVsCons}
	}
	\caption{The behavior of DWBAP limiter in terms of conservative variables and characteristic variables. Char. denotes the characteristic variables. Cons. represents the conservative variables.}
	\label{fig:sodCharVsCons}
\end{figure}

The effects of limiters are also investigated. The accuracy of DWBAP limiter has been proved in the reference \cite{liu2017accuracy}. In this case, we only give some auxiliary notes. The comparison of DWBAP \cite{liu2017accuracy} and WBAP-L2 \cite{li2012multi} in terms of conservative variables is shown in Fig.~\ref{fig:sodLimiter}, and the results are obtained on a Cartesian grid. In general speaking, DWBAP limiter is a little more accurate than WBAP-L2 limiter. But, for suppressing oscillation, the WBAP-L2 limiter behaviors better. The effects of limiters in terms of conservative and characteristic variables are also considered on unstructured grid in Fig.~\ref{fig:sod:unstructured2}. Fig.~\ref{fig:sodCharVsCons} shows the difference between the DWBAP limiter in terms of conservative variables and characteristic variables respectively. It is obvious that the limiter in terms of characteristic variables has less oscillation. The behavior of high order limiter is a open question, and it is worthy of great efforts. The results above are not the final conclusion, and more investigations are under considered.

\subsection{Lax problem}
Lax problem is another one-dimensional Riemann problem. Compared with Sod problem, Lax problem has a more stronger discontinuity.
The initial condition is expressed as
\begin{equation}\label{initLax}
	\left(\rho, u, v, p\right)= \left\{
	\begin{array}{ll}
	\left(0.445, 0.698, 0, 3.528\right), & 0<x<0.5,\\
	\left(0.5, 0, 0, 0.571\right), & 0.5 \leq x \leq 1.
	\end{array}
	\right.
\end{equation}

\begin{figure}[!htp]
	\centering
	\subfigure[$\rho$]{
		\includegraphics[width=0.3 \textwidth]{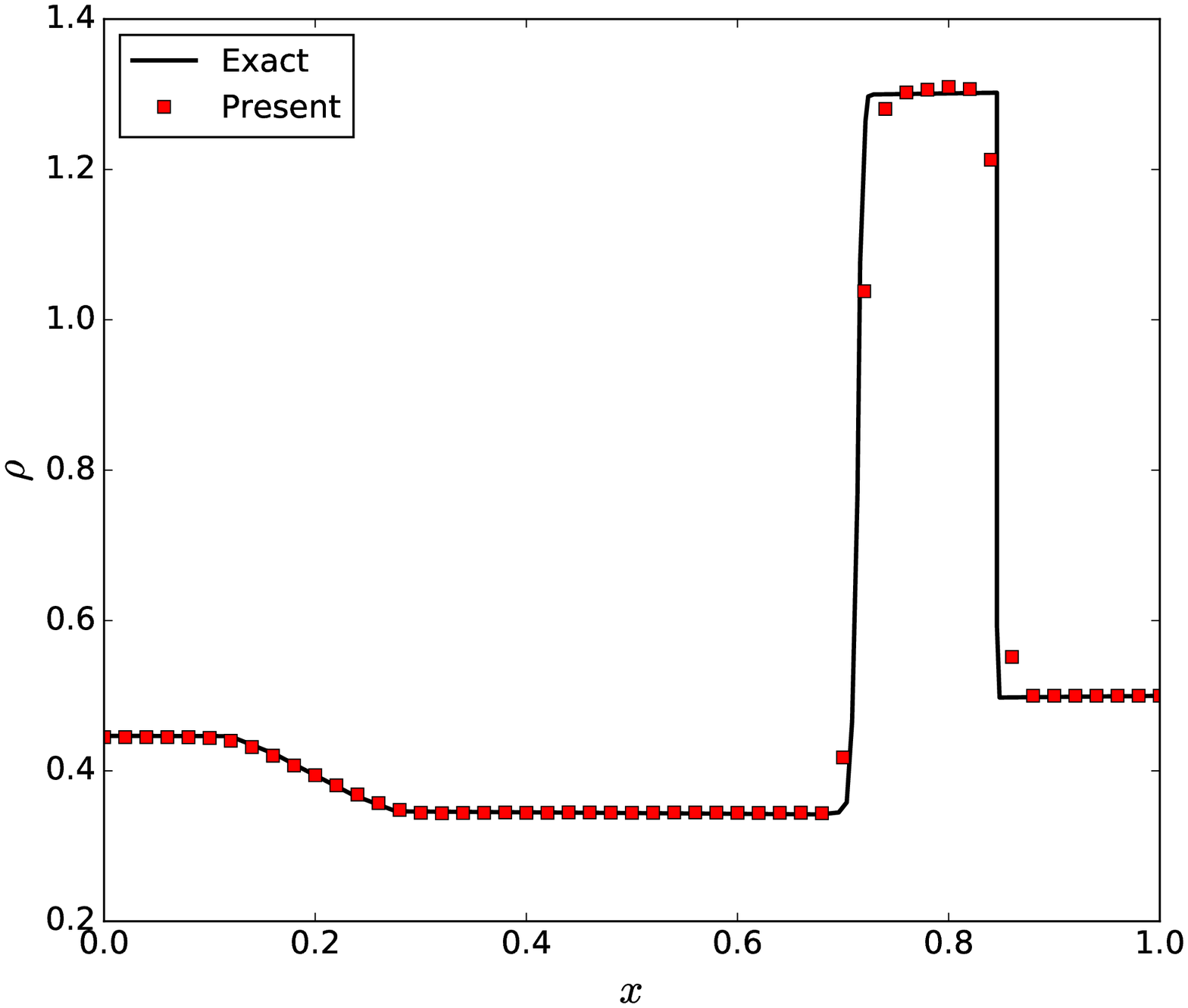}
		\label{fig:lax:den}
	}
	\subfigure[$u$]{
		\includegraphics[width=0.3 \textwidth]{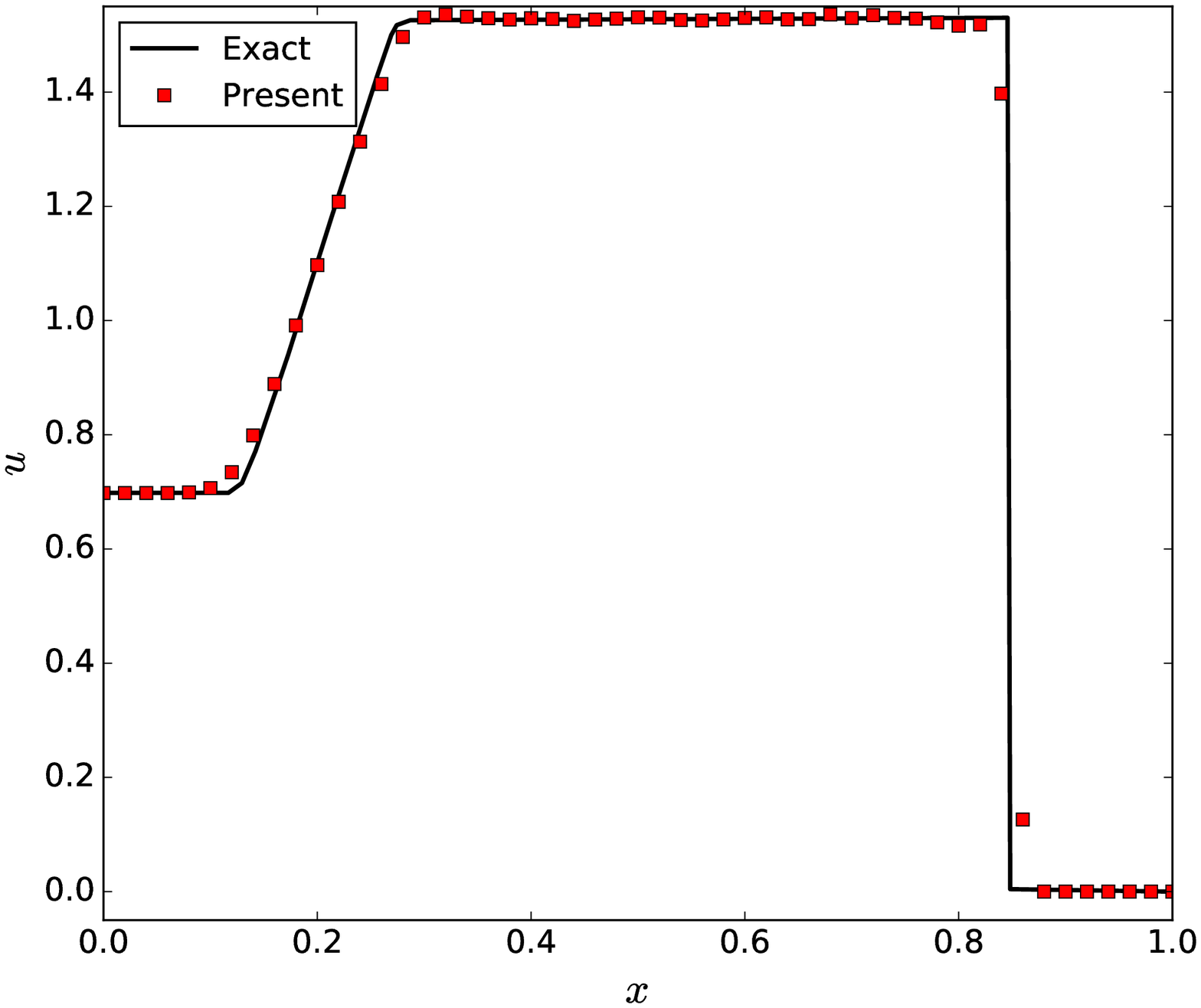}
		\label{fig:lax:u}
	}
	\subfigure[$p$]{
		\includegraphics[width=0.3 \textwidth]{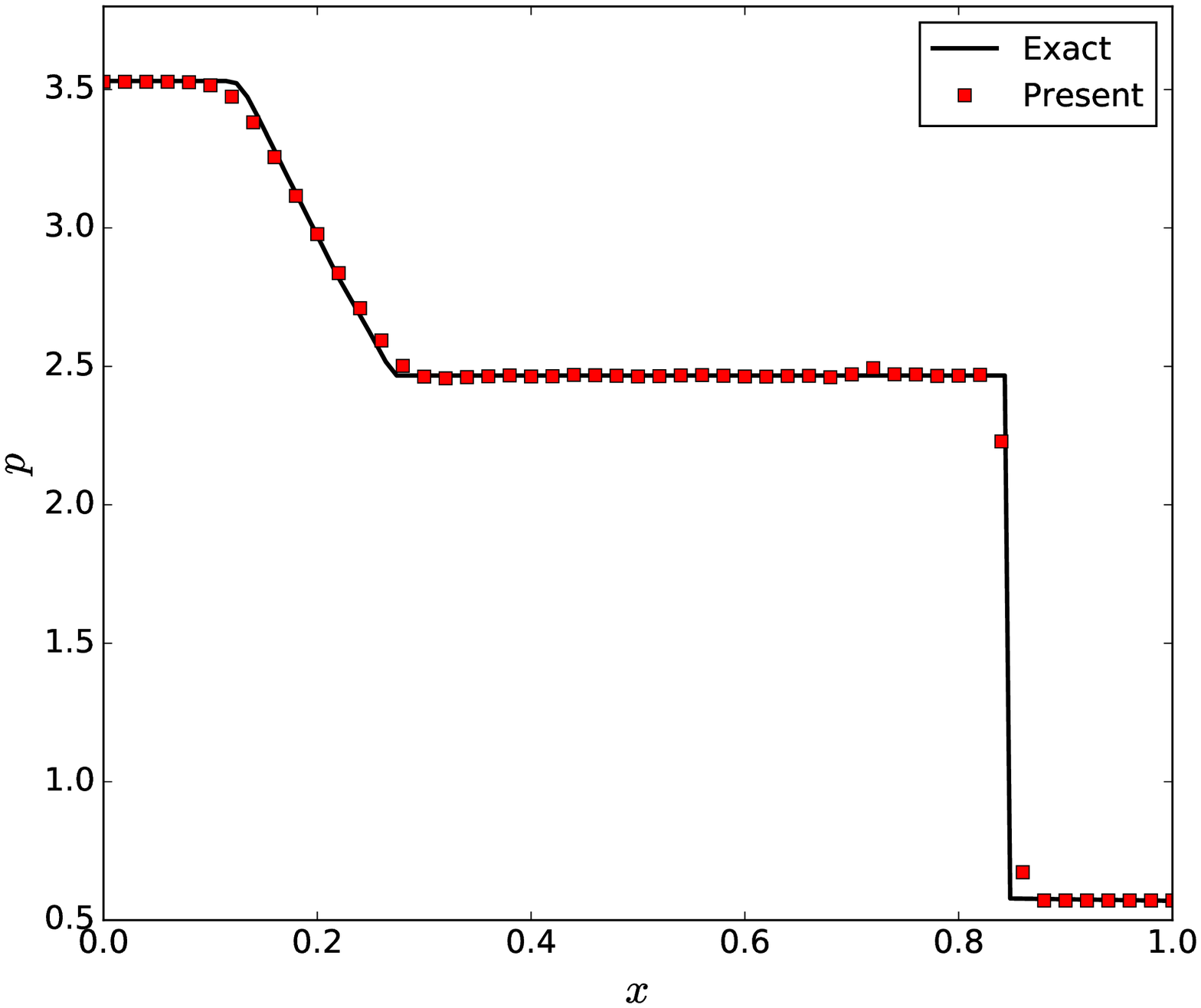}
		\label{fig:lax:p}
	}
	\caption{The numerical results of Lax problem.}
	\label{fig:lax}
\end{figure}

\begin{figure}[!htp]
	\centering
	\includegraphics[width=0.8 \textwidth]{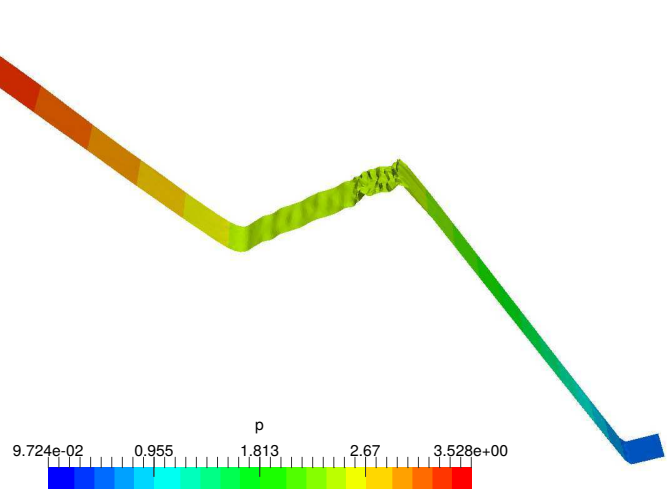}
	\caption{The three-dimensional view of pressure distribution.}
	\label{fig:lax:3Dp}
\end{figure}

The unstructured grid used in the computation is similar to the grid shown in Fig.~\ref{fig:sod:unstructured2}, and the length scale $h$ equals to $1/100$. The numerical results are obtained at the time $t=0.14$. Fig.~\ref{fig:lax} shows the distribution of density, velocity and pressure. The result computed using present method has a good agreement with the exact solution, and the discontinuities are captured accurately. Fig.~\ref{fig:lax:3Dp} shows the three-dimensional view of the pressure distribution. Because of the unstructured grid used in the computation, the small oscillation can be seen in the plot. Using the two-dimensional codes to simulate one-dimensional problem on a triangular grid, such a small oscillation is always existed. Not only is in Lax problem, but also in Sod problem and so on.

\subsection{Acoustic pressure pulse}
\begin{figure}[!htp]
	\centering
	\subfigure[Global view]{
		\includegraphics[width=0.4 \textwidth]{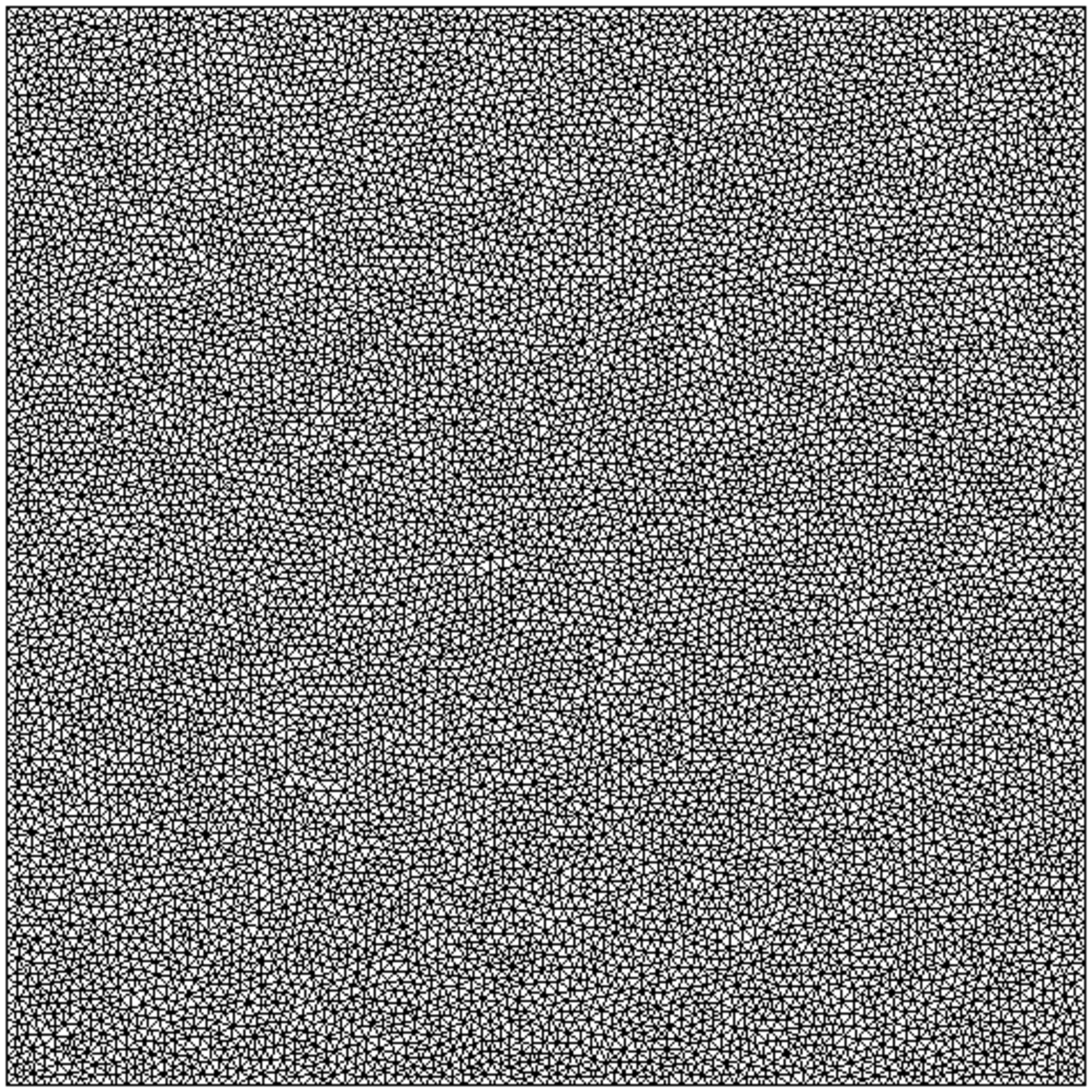}
		\label{fig:cirPulseGrid:global}
	}
	\subfigure[Zoom in view]{
		\includegraphics[width=0.4 \textwidth]{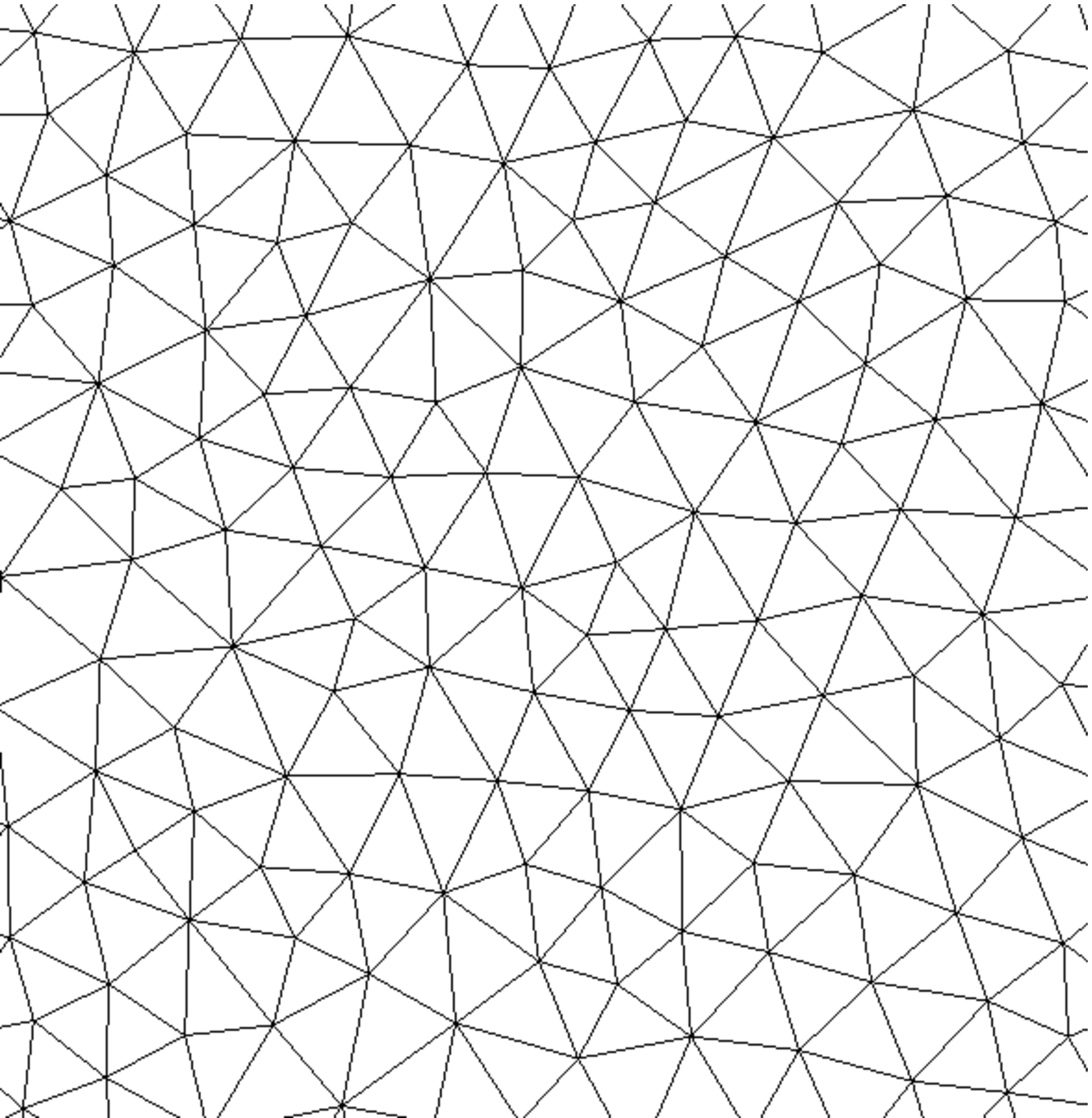}
		\label{fig:cirPulseGrid:zoom}
	}
	\caption{The unstructured grid ($h=1/80$) used in the simulation of acoustic pressure pulse problem.}
	\label{fig:cirPulseGrid}
\end{figure}

Two-dimensional acoustic pressure pulse problem is a case which the high order method has advantages in resolving the acoustic wave.
The initial perturbation is given by a Gaussian pressure distribution at the center of the computational domain at $t=0$.
\begin{equation}\label{InitPulse}
    \rho = \rho_\infty,\quad u=v=0,\quad p = p_\infty+\varepsilon e^{-\alpha\eta^2},
\end{equation}
where $\varepsilon=0.01$, $\eta=\sqrt{(x-0.5)^2+(y-0.5)^2}$ and $\alpha=\ln{2}/0.04^2$. The reference parameters are $p_{ref}=p_\infty$, $\rho_{ref}=\rho_\infty$, $u_{ref}=\sqrt{p_{ref}/\rho_{ref}}$ and $t_{ref}=l_{ref}/u_{ref}$. The analytical solution \cite{tam1993dispersion} at time $t$ can be given as

\begin{equation}\label{cirPulseSolution}
p = p_\infty+\frac{\epsilon}{2\alpha}\int_{0}^{\infty}e^{-\xi^2/(4\alpha)}cos(c_s\xi t)J_0(\xi\eta)\xi d\xi,
\end{equation}
where $c_s$ represents the sound speed. The computational domain is $[0, 1]\times[0,1]$. The unstructured grid used in the computation is exhibited in Fig.~\ref{fig:cirPulseGrid}, and the structured girds used are the Cartesian grids with different length scale $h$.

\begin{figure}[!htp]
	\centering
	\includegraphics[width=0.4 \textwidth]{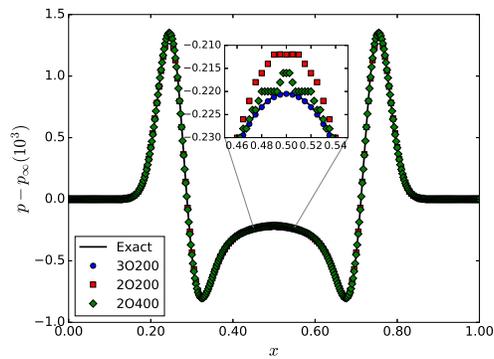}
	\caption{The comparison of the results between second and third order scheme.}
	\label{fig:cirPulse:2c3}
\end{figure}

\begin{figure}[!htp]
	\centering
	\includegraphics[width=0.4 \textwidth]{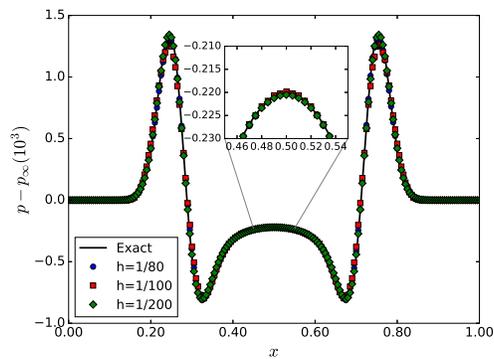}
	\caption{The results of Cartesian grid and triangular gird. The result labeled as $h=1/80$ is computed using triangular gird. The others are simulated on Cartesian grid.}
	\label{fig:cirPulse:TriVsCartesian}
\end{figure}

Fig.~\ref{fig:cirPulse:2c3} shows the numerical results of second and third order scheme. The legend `xOy' in picture means the simulation using xth order scheme on a Cartesian grid with $h=1/y$. A conclusion can be drawn that the third order scheme can approach the exact solution more accurate with fewer grid cells. The Fig.~\ref{fig:cirPulse:TriVsCartesian} exhibits the comparison of results using Cartesian grid and triangular grid, and all the results have good accordance with the analytical solution.

\subsection{Shu-Osher problem}
\begin{figure}[!htp]
	\centering
	\includegraphics[width=0.8 \textwidth]{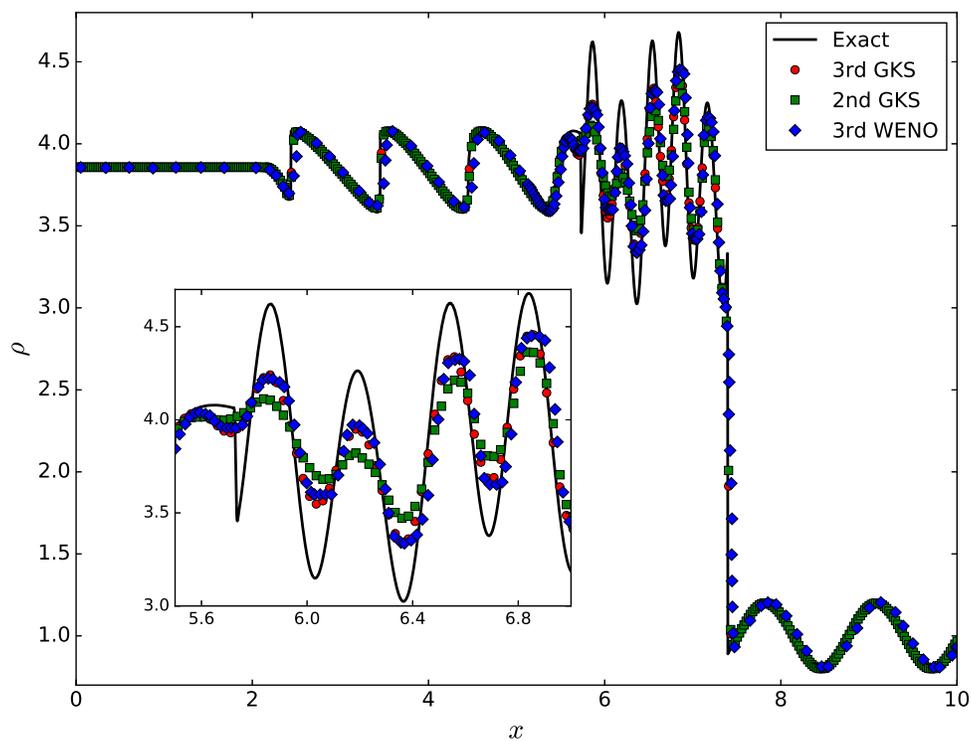}
	\caption{The density distribution of Shu-Osher problem.}
	\label{fig:shuOsher:rho}
\end{figure}

The problem of Shu-Osher \cite{shu1988efficient} describes the interaction of an entropy sin wave with a Mach 3 normal shock. The computational domain used in the simulation is taken as $[0, 10]\times[0,1]$. The initial condition is given as
\begin{equation}\label{initShuOsher}
	\left(\rho, u, v, p\right)=\left\{
		\begin{array}{ll}
			\left(3.857143, 2.629369, 0. 10.33333\right), & 0 \leq x < 1,\\
			\left(1+0.2sin(5x), 0, 0, 1\right), & 1 \leq x \leq 10.
		\end{array}
	\right.
\end{equation}
The numerical result is obtained at time $t=1.8$. In this case, the Cartesian grid is used, and the length scale $h$ is $1/80$. Because the exact solution of this problem can not be computed directly, the solution of fourth order WENO method with $10000$ grid in one dimension is taken as the exact result. The numerical results are shown in Fig.~\ref{fig:shuOsher:rho}. Compared to the numerical results of second order gas-kinetic scheme and third order WENO \cite{wang2015accurate}, the advantages of third order method in present work can be seen obviously.

\subsection{Shock-vortex interaction}
\begin{figure}[!htp]
	\centering
	\includegraphics[width=0.8 \textwidth]{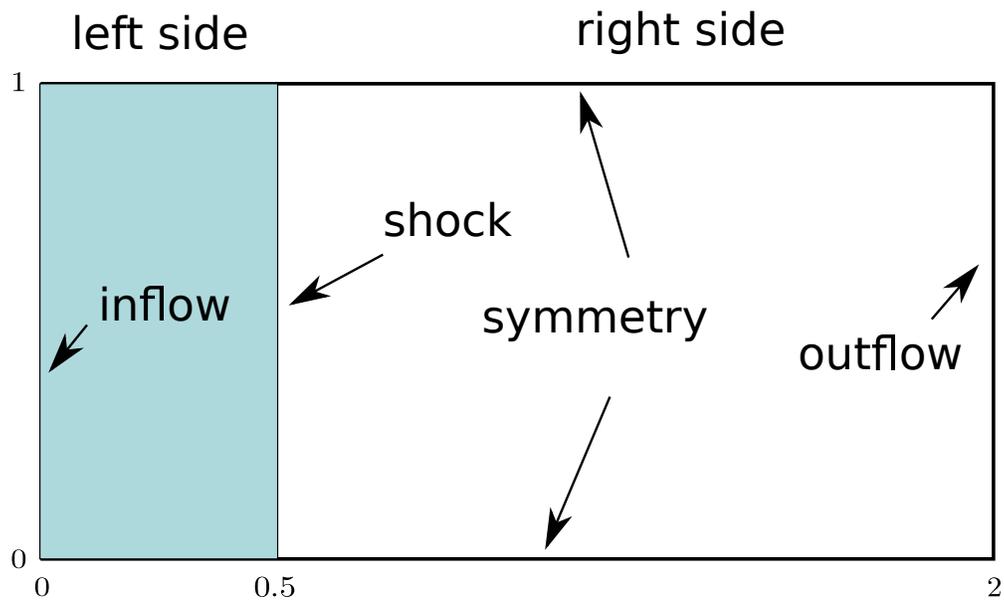}
	\caption{The computational domain and the boundary conditions in the simulation of shock-vortex interaction problem.}
	\label{fig:shockvortex:domain}
\end{figure}

\begin{figure}[!htp]
	\centering
	\subfigure[Global view]{
		\includegraphics[width=0.8 \textwidth]{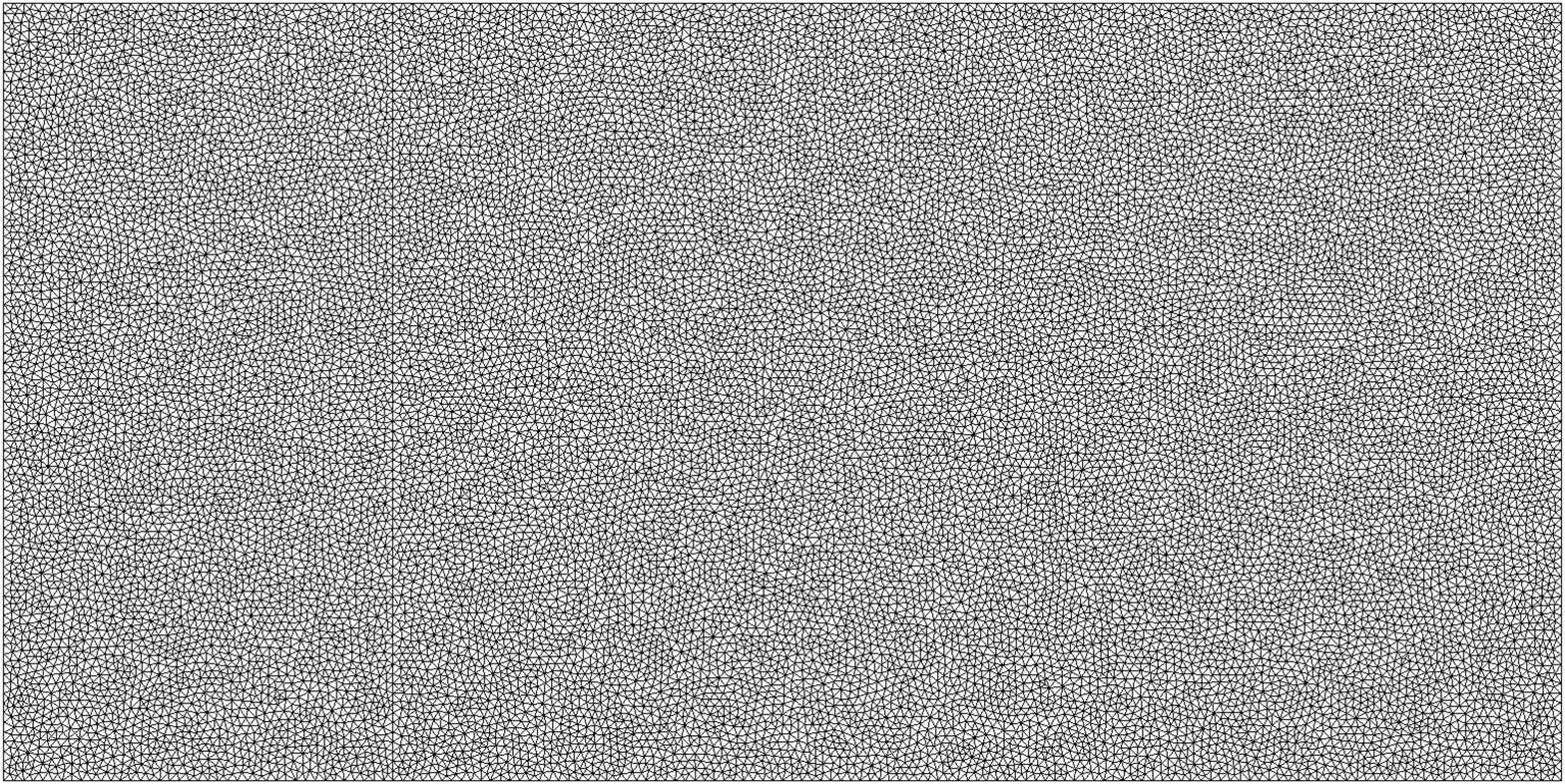}
		\label{fig:shockvortex:gridglobal}
	}
	\subfigure[Zoom in view]{
		\includegraphics[width=0.8 \textwidth]{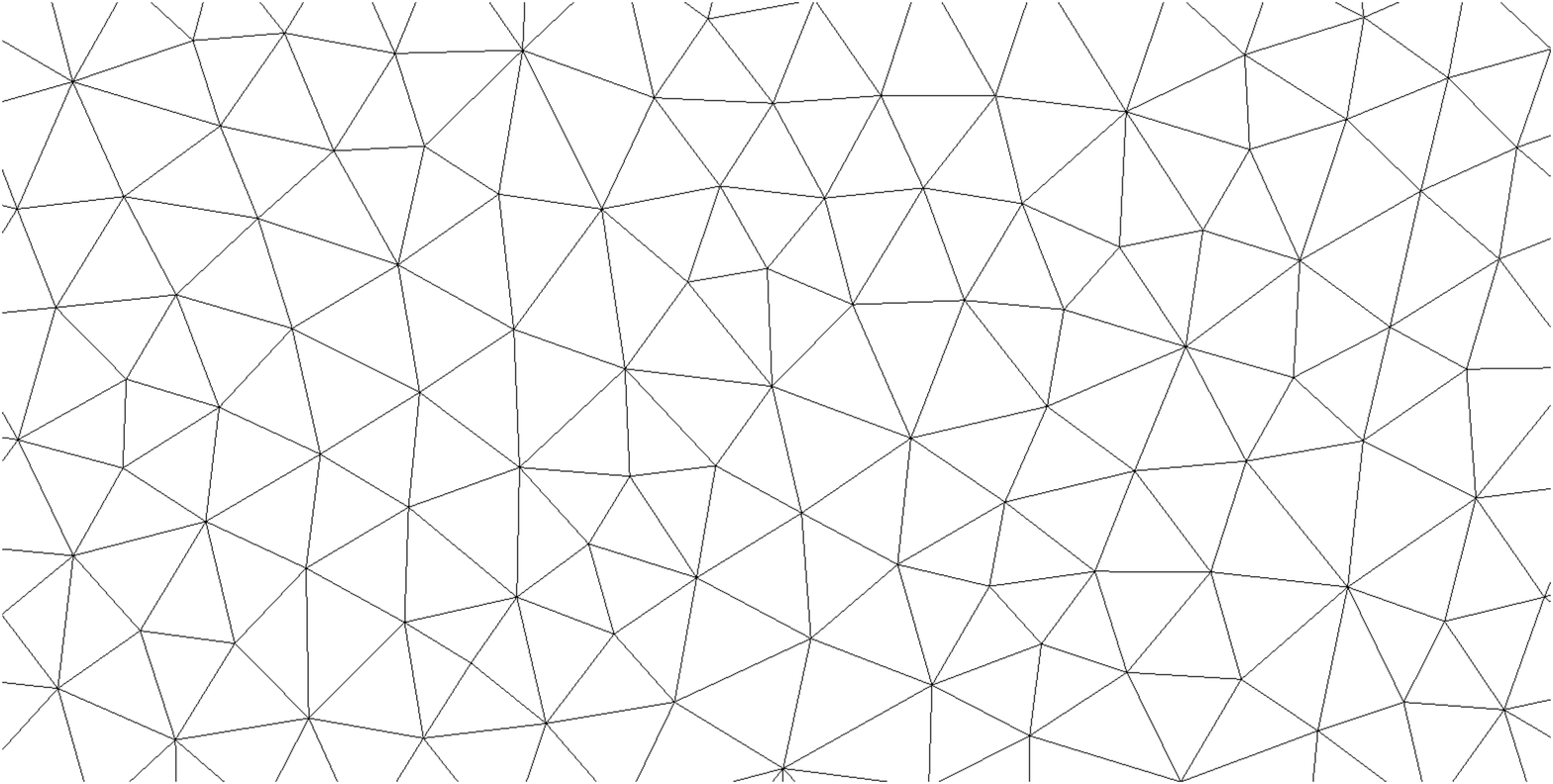}
		\label{fig:shockvortex:gridzoom}
	}
	\caption{The unstructured grid used in the simulation of shock-vortex interaction problem.}
	\label{fig:shockvortex:grid}
\end{figure}

\begin{figure}[!htp]
	\centering
	\subfigure[$t=0$]{
		\includegraphics[width=0.4 \textwidth]{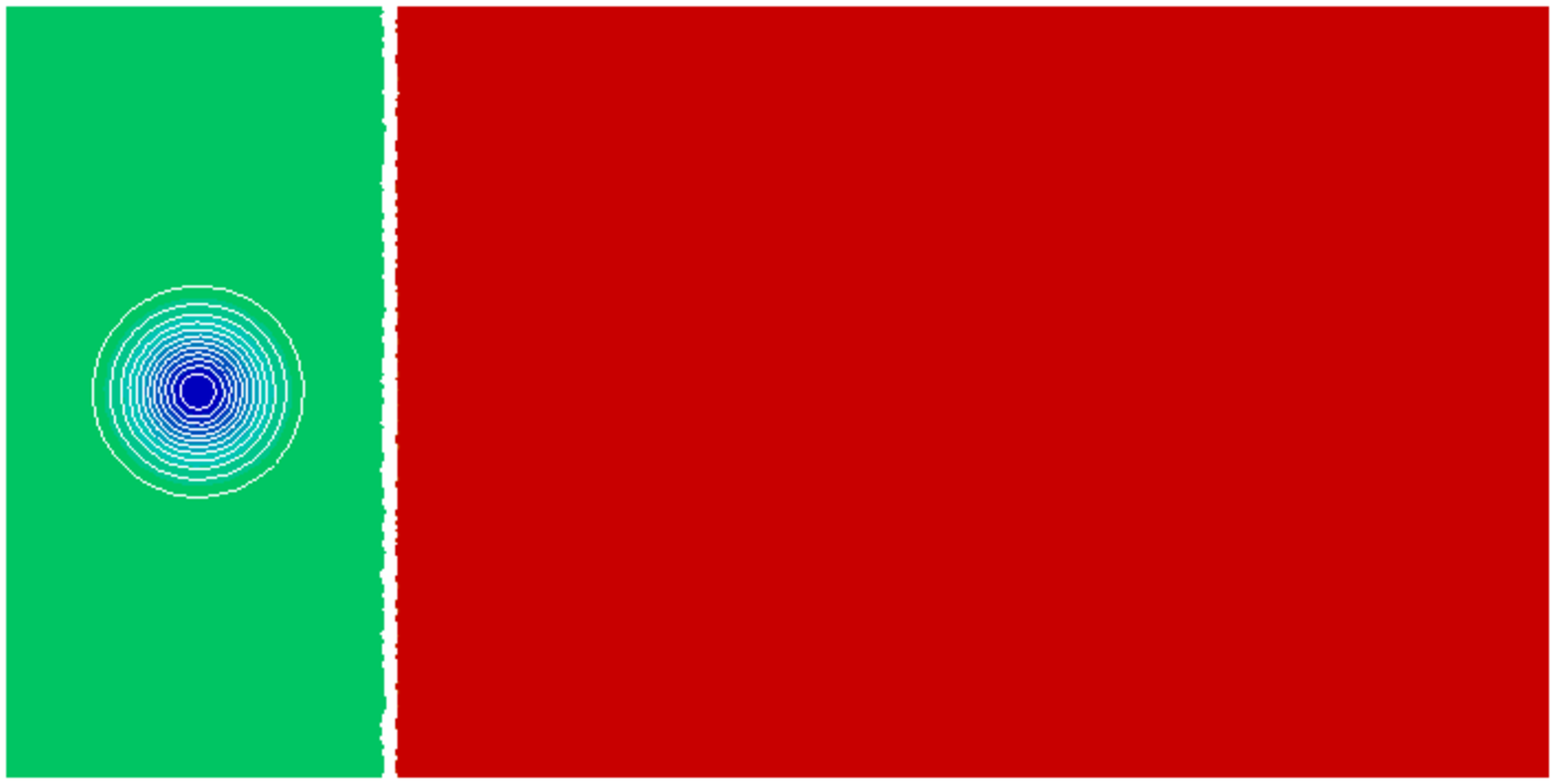}
		\label{fig:shockvortex:0}
	}
	\subfigure[$t=0.05$]{
		\includegraphics[width=0.4 \textwidth]{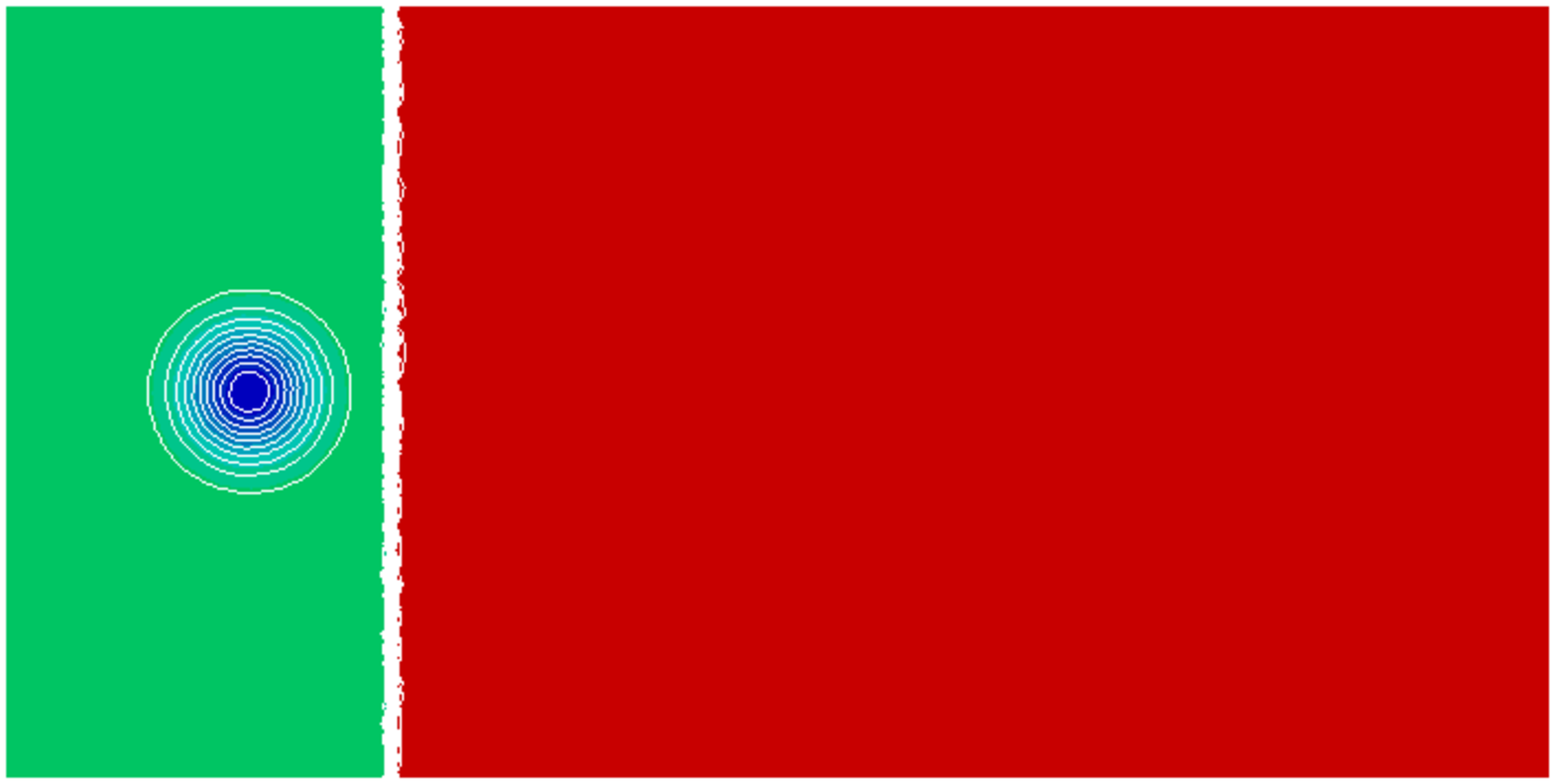}
		\label{fig:shockvortex:1}
	}
	\subfigure[$t=0.2$]{
		\includegraphics[width=0.4 \textwidth]{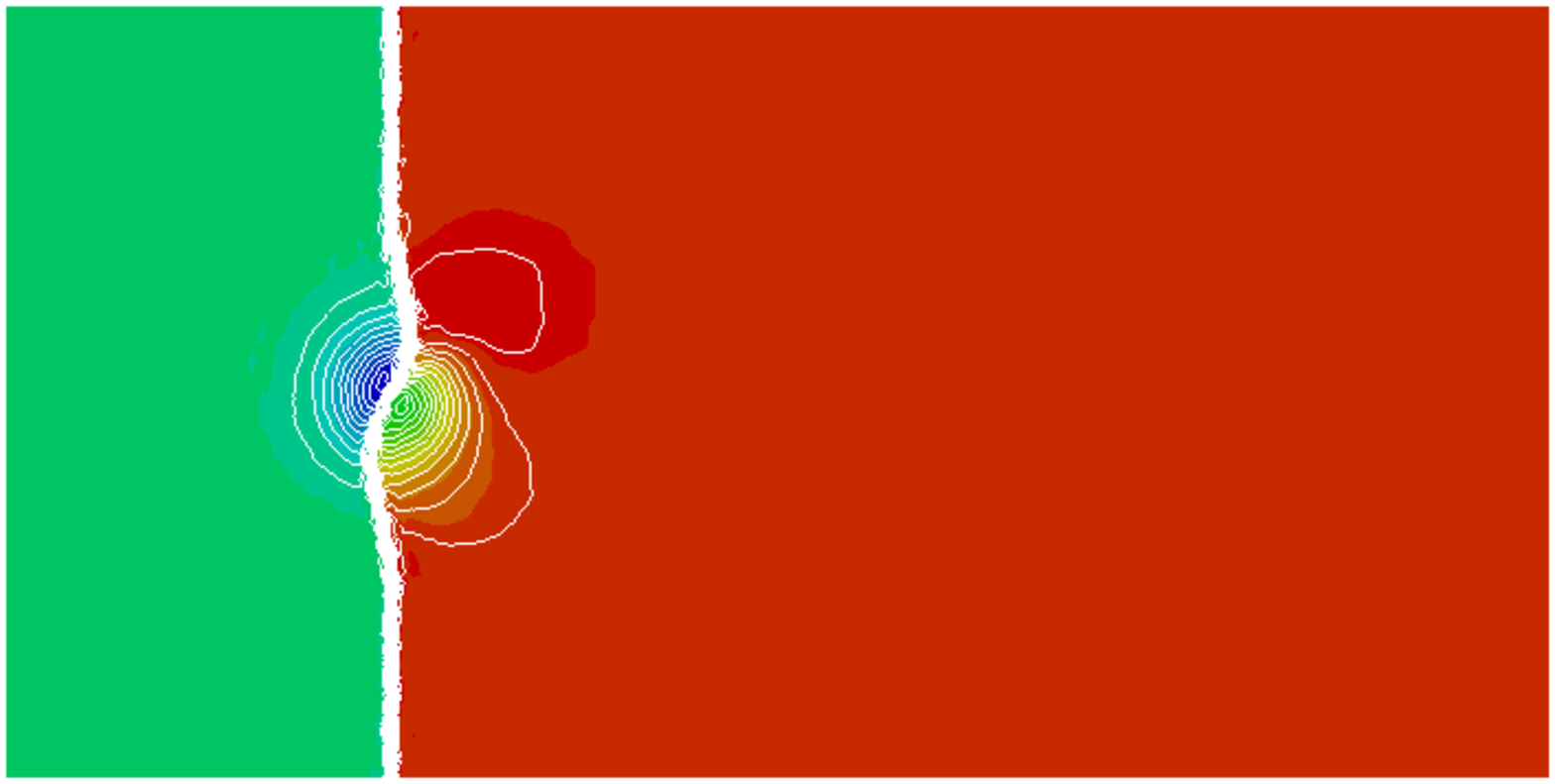}
		\label{fig:shockvortex:2}
	}
	\subfigure[$t=0.35$]{
		\includegraphics[width=0.4 \textwidth]{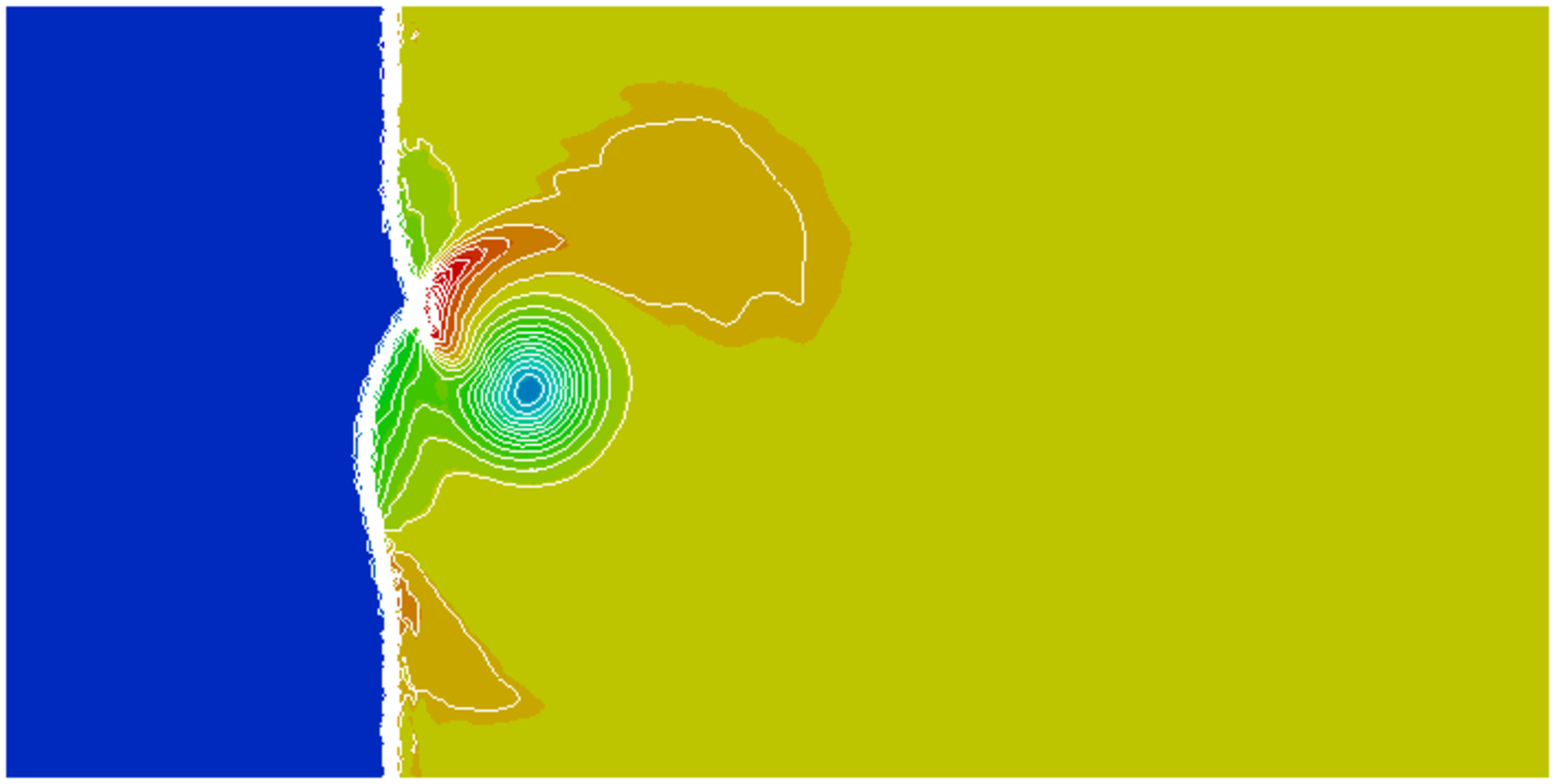}
		\label{fig:shockvortex:3}
	}
	\caption{The pressure distribution of the shock-vortex interaction problem.}
	\label{fig:shockvortex}
\end{figure}

\begin{figure}[!htp]
	\centering
	\includegraphics[width=0.8 \textwidth]{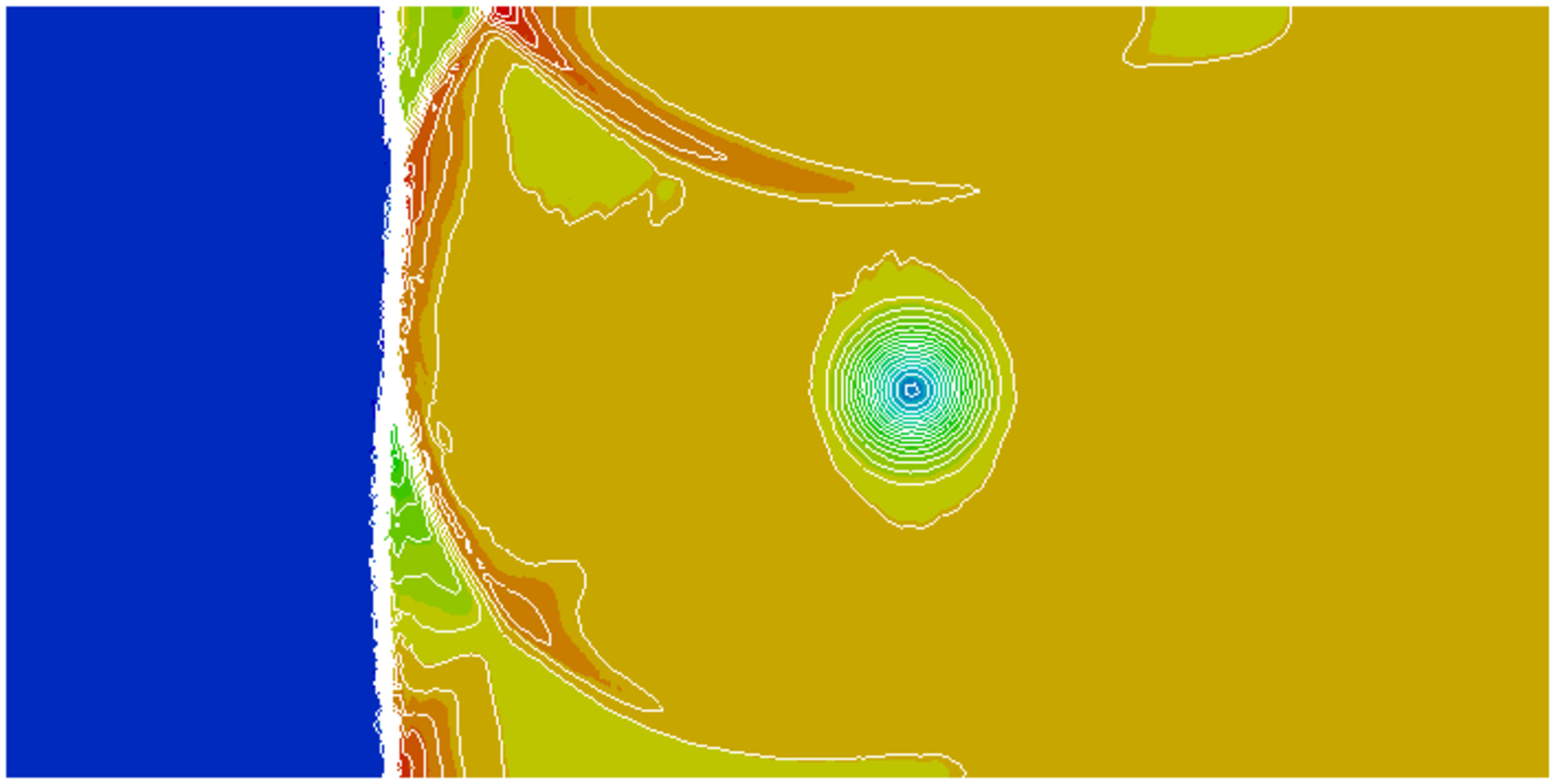}
	\caption{The pressure distribution of the shock-vortex interaction problem at $t=0.8$.}
	\label{fig:shockvortex:4}
\end{figure}

High order methods have some advantages in the problem of shock-vortex interaction \cite{shu1998essentially}, as it resolves the vortex and the interaction better.
In the simulation, a stationary normal shock is initialed in the flow field. A Mach 1.1 normal shock wave is located at the position $x=0.5$. The left side state ($Ma=1.1$) of the shock wave is given as follows
\begin{equation}\label{shockVortexLeft}
		(\rho,u,v,p)=(1.0, Ma\sqrt{\gamma}, 0, 1.0),\quad T=p/\rho,\quad S=ln(p/\rho^{\gamma}).
\end{equation}
A small and weak vortex is superposed to the left side of the normal shock. The center of the vortex is $(x_c, y_c)=(0.25, 0.5)$. The perturbation is given as
\begin{equation}\label{shockVortexPerturbation1}
	(\delta u, \delta v) = \kappa\eta e^{\mu(1-\eta^2)}(sin\theta,-cos\theta),
\end{equation}
\begin{equation}\label{shockVortexPerturbation2}
	\delta T = -\frac{(\gamma-1)\kappa^2}{4\mu\gamma}e^{2\mu(1-\eta^2)}(sin\theta,-cos\theta), \delta S=0,
\end{equation}
where
\begin{equation}\label{shockvortexcoefficients}
	\begin{aligned}
	\kappa &= 0.3, \\
     \mu &= 0.204, \\
    \eta &= r/r_c, \\
    r_c &= 0.05, \\
    r &= \sqrt{(x-x_c)^2+(y-y_c)^2}.
	\end{aligned}
\end{equation}
The computational domain and the boundary conditions are shown in Fig.~\ref{fig:shockvortex:domain}. The unstructured grid shown in Fig.~\ref{fig:shockvortex:grid} is used in this case. The length scale $h$ is $1/80$. The numerical results shown in Fig.~\ref{fig:shockvortex} are the contours of pressure at different times, and it appears that our present method can capture the shock and the interaction with enough resolution. The picture exhibited in Fig.~\ref{fig:shockvortex:4} is the result at time $t=0.8$. It shows clearly that the shock bifurcations reaches to the top boundary, and the reflection is evident.

\subsection{Lid-driven cavity flow}
\begin{figure}[!htp]
	\centering
	\subfigure[$u$-velocity along the vertical central line.]{
		\includegraphics[width=0.4 \textwidth]{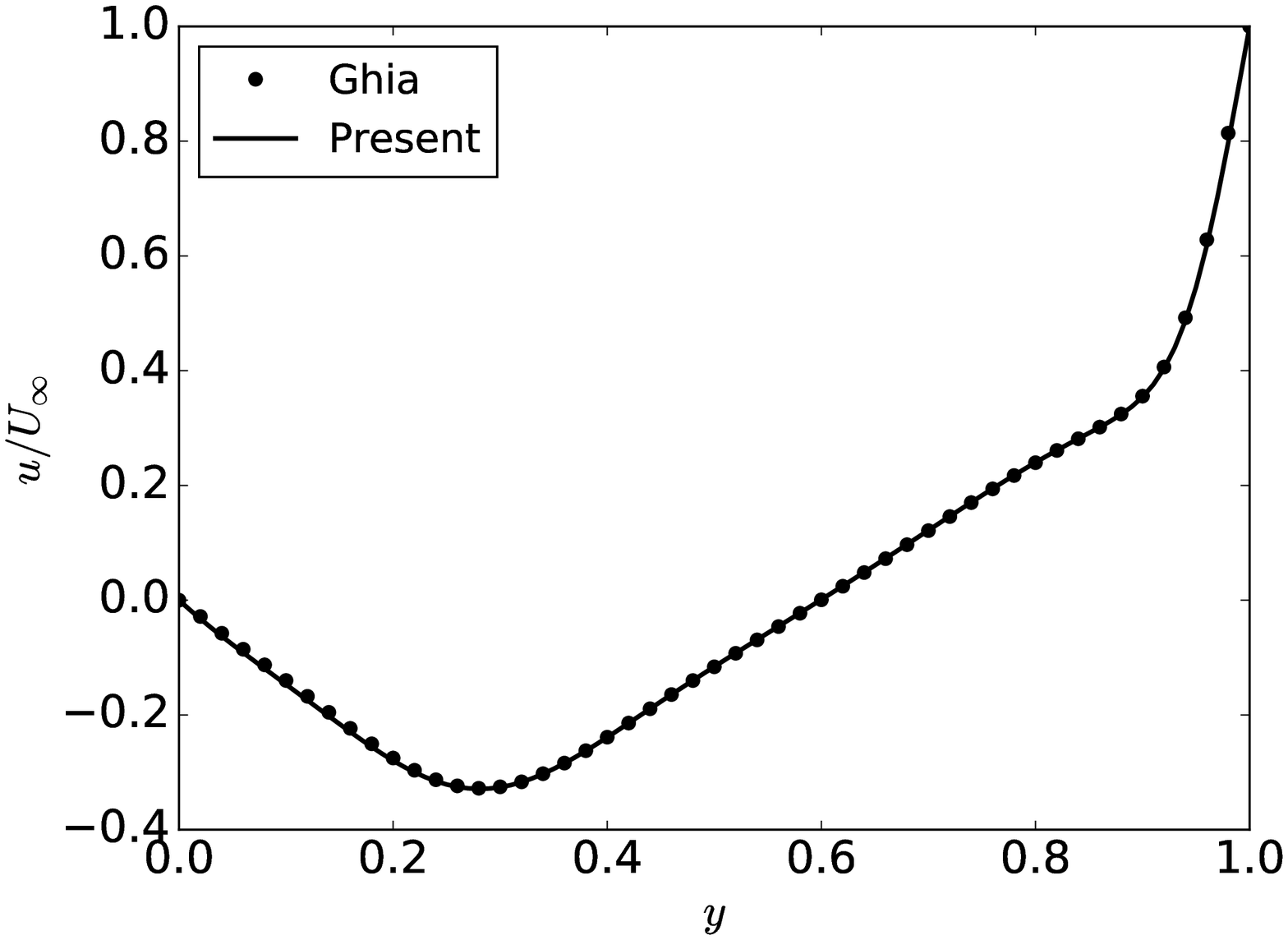}
		\label{fig:cavity400:u}
	}
	\subfigure[$v$-velocity along the horizontal central line.]{
		\includegraphics[width=0.4 \textwidth]{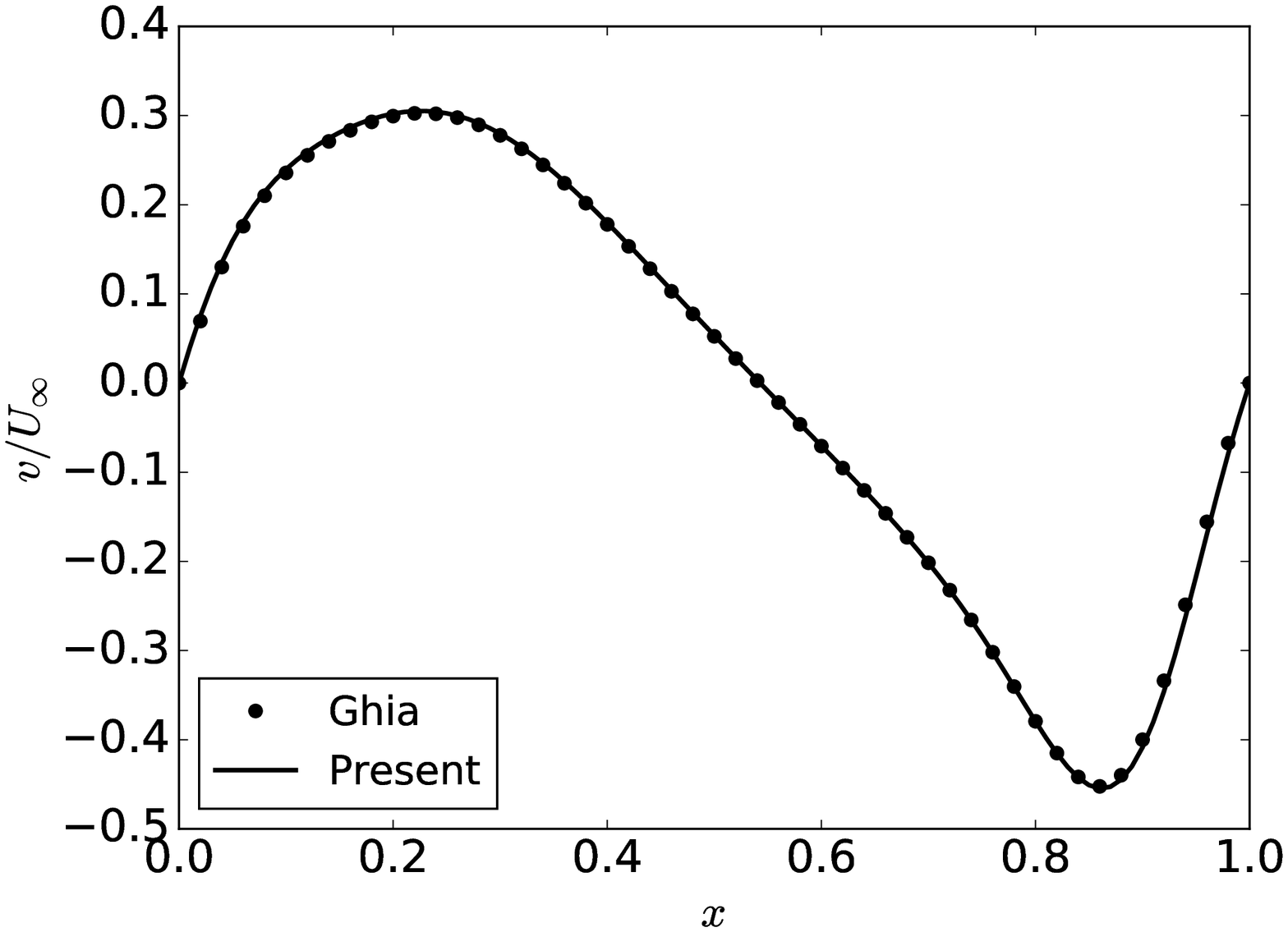}
		\label{fig:cavity400:v}
	}
	\caption{The velocity profiles of cavity flow at $Re=400$.}
	\label{fig:cavity400}
\end{figure}

\begin{figure}[!htp]
	\centering
	\subfigure[$u$-velocity along the vertical central line.]{
		\includegraphics[width=0.4 \textwidth]{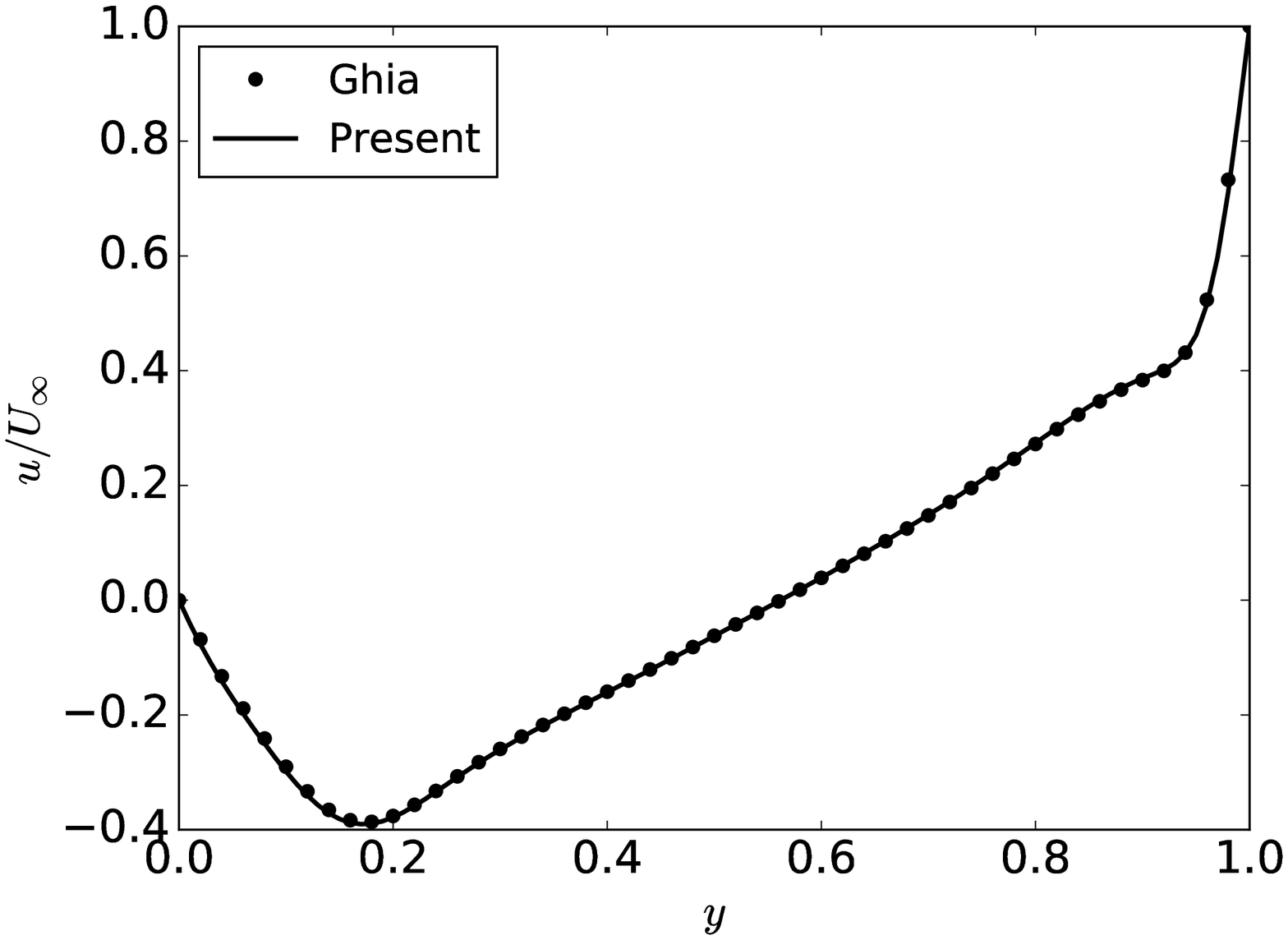}
		\label{fig:cavity1000:u}
	}
	\subfigure[$v$-velocity along the horizontal central line.]{
		\includegraphics[width=0.4 \textwidth]{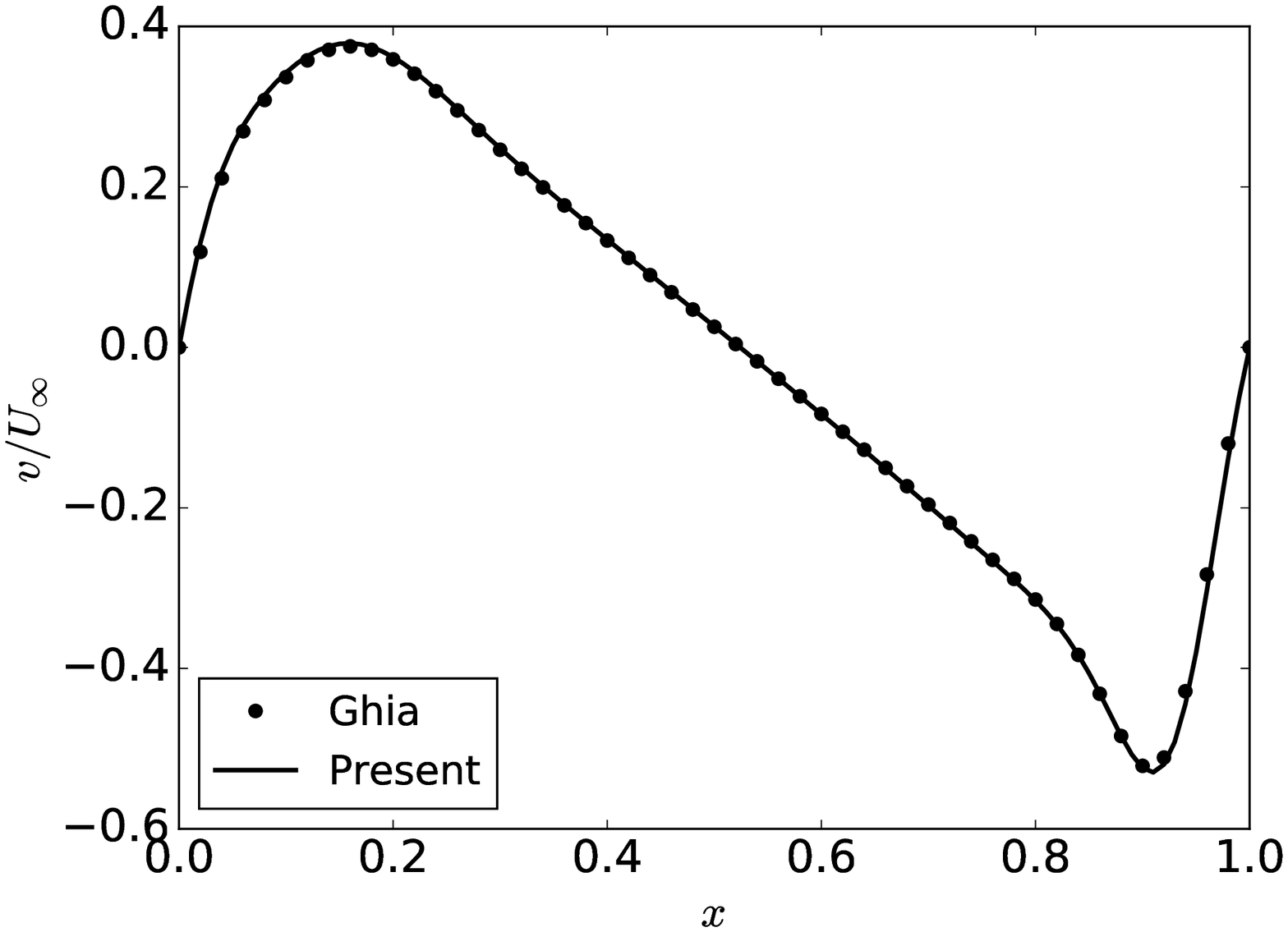}
		\label{fig:cavity1000:v}
	}
	\caption{The velocity profiles of cavity flow at $Re=1000$.}
	\label{fig:cavity1000}
\end{figure}

The lid-driven cavity flow is one of the benchmarks for validating the performance of the viscous flow solver. An incompressible flow is initialed in the computational domain, and the Mach number of the lid is set as $Ma=0.1$. The Reynolds number are $Re=400,1000$. The computational domain is $[0,1]\times[0,1]$, and the grid used in this case is the Cartesian grid with $h=1/65$. Fig.~\ref{fig:cavity400} and Fig.~\ref{fig:cavity1000} give numerical results at $Re=400, 1000$ respectively, and the results of present work have good accordance with the benchmark data of Ghia \cite{ghia1982high}. To reach such a good agreement with the benchmark data, the lower order method should use more grid cells to resolve the flow field compared with present method. It is evident that the numerical method proposed in our paper is also of enough accuracy for viscous flows.

\section{Conclusion}\label{conclusion}
In present paper, a high order gas-kinetic scheme is proposed based on the CLS reconstruction method. The compacted CLS method used in GKS can achieve third order accuracy on both unstructured and structured grids, which makes CLS method as a very promised algorithm in the field of high order finite volume method. Allied with the third order gas-kinetic flux solver, the third order scheme proposed in our work takes on both the advantages of gas-kinetic scheme and CLS algorithm. The numerical results exhibit that the accuracy designed is approached. Both the results using unstructured grid or structured grid are keeping good accordance with the benchmark data. To suppress oscillations near the discontinuity,  two kinds of limiters are investigated in the simulation, and some suggestions are given depended on the numerical results. The effects of limiter in term of conservative variables or characteristic variables are also considered in our work. Compared with the second order gas-kinetic scheme, present method can reach the same level accuracy using fewer grid cells.

\begin{acknowledgments}
The work has been financially supported by the National Natural Science Foundation of China (Grant No. 11472219), the 111 Project of China (B17037), as well as the ATCFD Project (2015-F-016).
\end{acknowledgments}

\clearpage
\section*{References}
\bibliography{ref}

 \newcommand{\noop}[1]{}
\begin{thebibliography}{45}%
\makeatletter
\providecommand \@ifxundefined [1]{%
 \@ifx{#1\undefined}
}%
\providecommand \@ifnum [1]{%
 \ifnum #1\expandafter \@firstoftwo
 \else \expandafter \@secondoftwo
 \fi
}%
\providecommand \@ifx [1]{%
 \ifx #1\expandafter \@firstoftwo
 \else \expandafter \@secondoftwo
 \fi
}%
\providecommand \natexlab [1]{#1}%
\providecommand \enquote  [1]{``#1''}%
\providecommand \bibnamefont  [1]{#1}%
\providecommand \bibfnamefont [1]{#1}%
\providecommand \citenamefont [1]{#1}%
\providecommand \href@noop [0]{\@secondoftwo}%
\providecommand \href [0]{\begingroup \@sanitize@url \@href}%
\providecommand \@href[1]{\@@startlink{#1}\@@href}%
\providecommand \@@href[1]{\endgroup#1\@@endlink}%
\providecommand \@sanitize@url [0]{\catcode `\\12\catcode `\$12\catcode
  `\&12\catcode `\#12\catcode `\^12\catcode `\_12\catcode `\%12\relax}%
\providecommand \@@startlink[1]{}%
\providecommand \@@endlink[0]{}%
\providecommand \url  [0]{\begingroup\@sanitize@url \@url }%
\providecommand \@url [1]{\endgroup\@href {#1}{\urlprefix }}%
\providecommand \urlprefix  [0]{URL }%
\providecommand \Eprint [0]{\href }%
\providecommand \doibase [0]{http://dx.doi.org/}%
\providecommand \selectlanguage [0]{\@gobble}%
\providecommand \bibinfo  [0]{\@secondoftwo}%
\providecommand \bibfield  [0]{\@secondoftwo}%
\providecommand \translation [1]{[#1]}%
\providecommand \BibitemOpen [0]{}%
\providecommand \bibitemStop [0]{}%
\providecommand \bibitemNoStop [0]{.\EOS\space}%
\providecommand \EOS [0]{\spacefactor3000\relax}%
\providecommand \BibitemShut  [1]{\csname bibitem#1\endcsname}%
\let\auto@bib@innerbib\@empty
\bibitem [{\citenamefont {Liao}\ \emph {et~al.}(2015)\citenamefont {Liao},
  \citenamefont {Ye},\ and\ \citenamefont {Zhang}}]{liao2015extending}%
  \BibitemOpen
  \bibfield  {author} {\bibinfo {author} {\bibfnamefont {F.}~\bibnamefont
  {Liao}}, \bibinfo {author} {\bibfnamefont {Z.}~\bibnamefont {Ye}}, \ and\
  \bibinfo {author} {\bibfnamefont {L.}~\bibnamefont {Zhang}},\ }\href@noop {}
  {\bibfield  {journal} {\bibinfo  {journal} {Journal of Computational
  Physics}\ }\textbf {\bibinfo {volume} {284}},\ \bibinfo {pages} {419}
  (\bibinfo {year} {2015})}\BibitemShut {NoStop}%
\bibitem [{\citenamefont {Barth}\ and\ \citenamefont
  {Frederickson}(1990)}]{barth1990higher}%
  \BibitemOpen
  \bibfield  {author} {\bibinfo {author} {\bibfnamefont {T.~J.}\ \bibnamefont
  {Barth}}\ and\ \bibinfo {author} {\bibfnamefont {P.~O.}\ \bibnamefont
  {Frederickson}},\ }\href@noop {} {\bibfield  {journal} {\bibinfo  {journal}
  {AIAA paper}\ }\textbf {\bibinfo {volume} {90}},\ \bibinfo {pages} {0013}
  (\bibinfo {year} {1990})}\BibitemShut {NoStop}%
\bibitem [{\citenamefont {Abgrall}(1994)}]{abgrall1994essentially}%
  \BibitemOpen
  \bibfield  {author} {\bibinfo {author} {\bibfnamefont {R.}~\bibnamefont
  {Abgrall}},\ }\href {\doibase https://doi.org/10.1006/jcph.1994.1148}
  {\bibfield  {journal} {\bibinfo  {journal} {Journal of Computational
  Physics}\ }\textbf {\bibinfo {volume} {114}},\ \bibinfo {pages} {45 }
  (\bibinfo {year} {1994})}\BibitemShut {NoStop}%
\bibitem [{\citenamefont {Durlofsky}\ \emph {et~al.}(1992)\citenamefont
  {Durlofsky}, \citenamefont {Engquist},\ and\ \citenamefont
  {Osher}}]{durlofsky1992triangle}%
  \BibitemOpen
  \bibfield  {author} {\bibinfo {author} {\bibfnamefont {L.~J.}\ \bibnamefont
  {Durlofsky}}, \bibinfo {author} {\bibfnamefont {B.}~\bibnamefont {Engquist}},
  \ and\ \bibinfo {author} {\bibfnamefont {S.}~\bibnamefont {Osher}},\ }\href
  {\doibase https://doi.org/10.1016/0021-9991(92)90173-V} {\bibfield  {journal}
  {\bibinfo  {journal} {Journal of Computational Physics}\ }\textbf {\bibinfo
  {volume} {98}},\ \bibinfo {pages} {64 } (\bibinfo {year} {1992})}\BibitemShut
  {NoStop}%
\bibitem [{\citenamefont {Ollivier-Gooch}(1997)}]{ollivier1997quasi}%
  \BibitemOpen
  \bibfield  {author} {\bibinfo {author} {\bibfnamefont {C.~F.}\ \bibnamefont
  {Ollivier-Gooch}},\ }\href {\doibase https://doi.org/10.1006/jcph.1996.5584}
  {\bibfield  {journal} {\bibinfo  {journal} {Journal of Computational
  Physics}\ }\textbf {\bibinfo {volume} {133}},\ \bibinfo {pages} {6 }
  (\bibinfo {year} {1997})}\BibitemShut {NoStop}%
\bibitem [{\citenamefont {Sonar}(1997)}]{sonar1997construction}%
  \BibitemOpen
  \bibfield  {author} {\bibinfo {author} {\bibfnamefont {T.}~\bibnamefont
  {Sonar}},\ }\href {\doibase https://doi.org/10.1016/S0045-7825(96)01060-2}
  {\bibfield  {journal} {\bibinfo  {journal} {Computer Methods in Applied
  Mechanics and Engineering}\ }\textbf {\bibinfo {volume} {140}},\ \bibinfo
  {pages} {157 } (\bibinfo {year} {1997})}\BibitemShut {NoStop}%
\bibitem [{\citenamefont {Liu}\ \emph {et~al.}(1994)\citenamefont {Liu},
  \citenamefont {Osher},\ and\ \citenamefont {Chan}}]{liu1994weighted}%
  \BibitemOpen
  \bibfield  {author} {\bibinfo {author} {\bibfnamefont {X.-D.}\ \bibnamefont
  {Liu}}, \bibinfo {author} {\bibfnamefont {S.}~\bibnamefont {Osher}}, \ and\
  \bibinfo {author} {\bibfnamefont {T.}~\bibnamefont {Chan}},\ }\href {\doibase
  https://doi.org/10.1006/jcph.1994.1187} {\bibfield  {journal} {\bibinfo
  {journal} {Journal of Computational Physics}\ }\textbf {\bibinfo {volume}
  {115}},\ \bibinfo {pages} {200 } (\bibinfo {year} {1994})}\BibitemShut
  {NoStop}%
\bibitem [{\citenamefont {Shu}(1998)}]{shu1998essentially}%
  \BibitemOpen
  \bibfield  {author} {\bibinfo {author} {\bibfnamefont {C.-W.}\ \bibnamefont
  {Shu}},\ }in\ \href@noop {} {\emph {\bibinfo {booktitle} {Advanced Numerical
  Approximation of Nonlinear Hyperbolic Equations}}}\ (\bibinfo  {publisher}
  {Springer},\ \bibinfo {year} {1998})\ pp.\ \bibinfo {pages}
  {325--432}\BibitemShut {NoStop}%
\bibitem [{\citenamefont {Hu}\ and\ \citenamefont
  {Shu}(1999)}]{hu1999weighted}%
  \BibitemOpen
  \bibfield  {author} {\bibinfo {author} {\bibfnamefont {C.}~\bibnamefont
  {Hu}}\ and\ \bibinfo {author} {\bibfnamefont {C.-W.}\ \bibnamefont {Shu}},\
  }\href {\doibase https://doi.org/10.1006/jcph.1998.6165} {\bibfield
  {journal} {\bibinfo  {journal} {Journal of Computational Physics}\ }\textbf
  {\bibinfo {volume} {150}},\ \bibinfo {pages} {97} (\bibinfo {year}
  {1999})}\BibitemShut {NoStop}%
\bibitem [{\citenamefont {Cockburn}\ and\ \citenamefont
  {Shu}(2001)}]{cockburn2001runge}%
  \BibitemOpen
  \bibfield  {author} {\bibinfo {author} {\bibfnamefont {B.}~\bibnamefont
  {Cockburn}}\ and\ \bibinfo {author} {\bibfnamefont {C.-W.}\ \bibnamefont
  {Shu}},\ }\href {\doibase 10.1023/A:1012873910884} {\bibfield  {journal}
  {\bibinfo  {journal} {Journal of Scientific Computing}\ }\textbf {\bibinfo
  {volume} {16}},\ \bibinfo {pages} {173} (\bibinfo {year} {2001})}\BibitemShut
  {NoStop}%
\bibitem [{\citenamefont {Liu}\ \emph {et~al.}(2016)\citenamefont {Liu},
  \citenamefont {Zhang}, \citenamefont {Jiang},\ and\ \citenamefont
  {Ye}}]{LIU20161096}%
  \BibitemOpen
  \bibfield  {author} {\bibinfo {author} {\bibfnamefont {Y.}~\bibnamefont
  {Liu}}, \bibinfo {author} {\bibfnamefont {W.}~\bibnamefont {Zhang}}, \bibinfo
  {author} {\bibfnamefont {Y.}~\bibnamefont {Jiang}}, \ and\ \bibinfo {author}
  {\bibfnamefont {Z.}~\bibnamefont {Ye}},\ }\href {\doibase
  https://doi.org/10.1016/j.camwa.2016.06.024} {\bibfield  {journal} {\bibinfo
  {journal} {Computers \& Mathematics with Applications}\ }\textbf {\bibinfo
  {volume} {72}},\ \bibinfo {pages} {1096 } (\bibinfo {year}
  {2016})}\BibitemShut {NoStop}%
\bibitem [{\citenamefont {Wang}(2007)}]{wang2007high}%
  \BibitemOpen
  \bibfield  {author} {\bibinfo {author} {\bibfnamefont {Z.}~\bibnamefont
  {Wang}},\ }\href {\doibase https://doi.org/10.1016/j.paerosci.2007.05.001}
  {\bibfield  {journal} {\bibinfo  {journal} {Progress in Aerospace Sciences}\
  }\textbf {\bibinfo {volume} {43}},\ \bibinfo {pages} {1} (\bibinfo {year}
  {2007})}\BibitemShut {NoStop}%
\bibitem [{\citenamefont {Bhatnagar}\ \emph {et~al.}(1954)\citenamefont
  {Bhatnagar}, \citenamefont {Gross},\ and\ \citenamefont
  {Krook}}]{bhatnagar1954model}%
  \BibitemOpen
  \bibfield  {author} {\bibinfo {author} {\bibfnamefont {P.~L.}\ \bibnamefont
  {Bhatnagar}}, \bibinfo {author} {\bibfnamefont {E.~P.}\ \bibnamefont
  {Gross}}, \ and\ \bibinfo {author} {\bibfnamefont {M.}~\bibnamefont
  {Krook}},\ }\href@noop {} {\bibfield  {journal} {\bibinfo  {journal}
  {Physical Review}\ }\textbf {\bibinfo {volume} {94}},\ \bibinfo {pages} {511}
  (\bibinfo {year} {1954})}\BibitemShut {NoStop}%
\bibitem [{\citenamefont {Xu}(2001)}]{xu2001gas}%
  \BibitemOpen
  \bibfield  {author} {\bibinfo {author} {\bibfnamefont {K.}~\bibnamefont
  {Xu}},\ }\href@noop {} {\bibfield  {journal} {\bibinfo  {journal} {Journal of
  Computational Physics}\ }\textbf {\bibinfo {volume} {171}},\ \bibinfo {pages}
  {289} (\bibinfo {year} {2001})}\BibitemShut {NoStop}%
\bibitem [{\citenamefont {Xu}\ \emph {et~al.}(2005)\citenamefont {Xu},
  \citenamefont {Mao},\ and\ \citenamefont {Tang}}]{xu2005multidimensional}%
  \BibitemOpen
  \bibfield  {author} {\bibinfo {author} {\bibfnamefont {K.}~\bibnamefont
  {Xu}}, \bibinfo {author} {\bibfnamefont {M.}~\bibnamefont {Mao}}, \ and\
  \bibinfo {author} {\bibfnamefont {L.}~\bibnamefont {Tang}},\ }\href@noop {}
  {\bibfield  {journal} {\bibinfo  {journal} {Journal of Computational
  Physics}\ }\textbf {\bibinfo {volume} {203}},\ \bibinfo {pages} {405}
  (\bibinfo {year} {2005})}\BibitemShut {NoStop}%
\bibitem [{\citenamefont {Xiong}\ \emph {et~al.}(2011)\citenamefont {Xiong},
  \citenamefont {Zhong}, \citenamefont {Zhuo}, \citenamefont {Li},
  \citenamefont {Chen},\ and\ \citenamefont {Cao}}]{xiong2011numerical}%
  \BibitemOpen
  \bibfield  {author} {\bibinfo {author} {\bibfnamefont {S.}~\bibnamefont
  {Xiong}}, \bibinfo {author} {\bibfnamefont {C.}~\bibnamefont {Zhong}},
  \bibinfo {author} {\bibfnamefont {C.}~\bibnamefont {Zhuo}}, \bibinfo {author}
  {\bibfnamefont {K.}~\bibnamefont {Li}}, \bibinfo {author} {\bibfnamefont
  {X.}~\bibnamefont {Chen}}, \ and\ \bibinfo {author} {\bibfnamefont
  {J.}~\bibnamefont {Cao}},\ }\href@noop {} {\bibfield  {journal} {\bibinfo
  {journal} {International Journal for Numerical Methods in Fluids}\ }\textbf
  {\bibinfo {volume} {67}},\ \bibinfo {pages} {1833} (\bibinfo {year}
  {2011})}\BibitemShut {NoStop}%
\bibitem [{\citenamefont {Pan}\ \emph {et~al.}(2016{\natexlab{a}})\citenamefont
  {Pan}, \citenamefont {Zhong}, \citenamefont {Li},\ and\ \citenamefont
  {Zhuo}}]{FLD:FLD4239}%
  \BibitemOpen
  \bibfield  {author} {\bibinfo {author} {\bibfnamefont {D.}~\bibnamefont
  {Pan}}, \bibinfo {author} {\bibfnamefont {C.}~\bibnamefont {Zhong}}, \bibinfo
  {author} {\bibfnamefont {J.}~\bibnamefont {Li}}, \ and\ \bibinfo {author}
  {\bibfnamefont {C.}~\bibnamefont {Zhuo}},\ }\href@noop {} {\bibfield
  {journal} {\bibinfo  {journal} {International Journal for Numerical Methods
  in Fluids}\ }\textbf {\bibinfo {volume} {82}},\ \bibinfo {pages} {748}
  (\bibinfo {year} {2016}{\natexlab{a}})}\BibitemShut {NoStop}%
\bibitem [{\citenamefont {Li}\ \emph {et~al.}(2016)\citenamefont {Li},
  \citenamefont {Zhong}, \citenamefont {Pan},\ and\ \citenamefont
  {Zhuo}}]{li2016gas}%
  \BibitemOpen
  \bibfield  {author} {\bibinfo {author} {\bibfnamefont {J.}~\bibnamefont
  {Li}}, \bibinfo {author} {\bibfnamefont {C.}~\bibnamefont {Zhong}}, \bibinfo
  {author} {\bibfnamefont {D.}~\bibnamefont {Pan}}, \ and\ \bibinfo {author}
  {\bibfnamefont {C.}~\bibnamefont {Zhuo}},\ }\href {\doibase
  https://doi.org/10.1016/j.camwa.2016.09.012} {\bibfield  {journal} {\bibinfo
  {journal} {Computers \& Mathematics with Applications}\ } (\bibinfo {year}
  {2016}),\ https://doi.org/10.1016/j.camwa.2016.09.012}\BibitemShut {NoStop}%
\bibitem [{\citenamefont {Li}\ \emph {et~al.}(2005)\citenamefont {Li},
  \citenamefont {Fu},\ and\ \citenamefont {Xu}}]{li2005application}%
  \BibitemOpen
  \bibfield  {author} {\bibinfo {author} {\bibfnamefont {Q.}~\bibnamefont
  {Li}}, \bibinfo {author} {\bibfnamefont {S.}~\bibnamefont {Fu}}, \ and\
  \bibinfo {author} {\bibfnamefont {K.}~\bibnamefont {Xu}},\ }\href {\doibase
  10.2514/1.14130} {\bibfield  {journal} {\bibinfo  {journal} {AIAA Journal}\
  }\textbf {\bibinfo {volume} {43}},\ \bibinfo {pages} {2170} (\bibinfo {year}
  {2005})}\BibitemShut {NoStop}%
\bibitem [{\citenamefont {Yuan}\ \emph {et~al.}(2015)\citenamefont {Yuan},
  \citenamefont {Zhong},\ and\ \citenamefont {Zhang}}]{yuan2015immersed}%
  \BibitemOpen
  \bibfield  {author} {\bibinfo {author} {\bibfnamefont {R.}~\bibnamefont
  {Yuan}}, \bibinfo {author} {\bibfnamefont {C.}~\bibnamefont {Zhong}}, \ and\
  \bibinfo {author} {\bibfnamefont {H.}~\bibnamefont {Zhang}},\ }\href@noop {}
  {\bibfield  {journal} {\bibinfo  {journal} {Journal of Computational
  Physics}\ }\textbf {\bibinfo {volume} {296}},\ \bibinfo {pages} {184}
  (\bibinfo {year} {2015})}\BibitemShut {NoStop}%
\bibitem [{\citenamefont {Yuan}\ and\ \citenamefont
  {Zhong}(2018)}]{YUAN2018417}%
  \BibitemOpen
  \bibfield  {author} {\bibinfo {author} {\bibfnamefont {R.}~\bibnamefont
  {Yuan}}\ and\ \bibinfo {author} {\bibfnamefont {C.}~\bibnamefont {Zhong}},\
  }\href {\doibase https://doi.org/10.1016/j.apm.2017.10.003} {\bibfield
  {journal} {\bibinfo  {journal} {Applied Mathematical Modelling}\ }\textbf
  {\bibinfo {volume} {55}},\ \bibinfo {pages} {417 } (\bibinfo {year}
  {2018})}\BibitemShut {NoStop}%
\bibitem [{\citenamefont {Li}\ \emph {et~al.}(2014)\citenamefont {Li},
  \citenamefont {Kaneda},\ and\ \citenamefont {Suga}}]{li2014implicit}%
  \BibitemOpen
  \bibfield  {author} {\bibinfo {author} {\bibfnamefont {W.}~\bibnamefont
  {Li}}, \bibinfo {author} {\bibfnamefont {M.}~\bibnamefont {Kaneda}}, \ and\
  \bibinfo {author} {\bibfnamefont {K.}~\bibnamefont {Suga}},\ }\href@noop {}
  {\bibfield  {journal} {\bibinfo  {journal} {Computers \& Fluids}\ }\textbf
  {\bibinfo {volume} {93}},\ \bibinfo {pages} {100} (\bibinfo {year}
  {2014})}\BibitemShut {NoStop}%
\bibitem [{\citenamefont {Li}\ \emph {et~al.}(2017)\citenamefont {Li},
  \citenamefont {Zhong}, \citenamefont {Wang},\ and\ \citenamefont
  {Zhuo}}]{PhysRevE.95.053307}%
  \BibitemOpen
  \bibfield  {author} {\bibinfo {author} {\bibfnamefont {J.}~\bibnamefont
  {Li}}, \bibinfo {author} {\bibfnamefont {C.}~\bibnamefont {Zhong}}, \bibinfo
  {author} {\bibfnamefont {Y.}~\bibnamefont {Wang}}, \ and\ \bibinfo {author}
  {\bibfnamefont {C.}~\bibnamefont {Zhuo}},\ }\href {\doibase
  10.1103/PhysRevE.95.053307} {\bibfield  {journal} {\bibinfo  {journal}
  {Physical Review E}\ }\textbf {\bibinfo {volume} {95}},\ \bibinfo {pages}
  {053307} (\bibinfo {year} {2017})}\BibitemShut {NoStop}%
\bibitem [{\citenamefont {Li}\ \emph {et~al.}(2010)\citenamefont {Li},
  \citenamefont {Xu},\ and\ \citenamefont {Fu}}]{li2010high}%
  \BibitemOpen
  \bibfield  {author} {\bibinfo {author} {\bibfnamefont {Q.}~\bibnamefont
  {Li}}, \bibinfo {author} {\bibfnamefont {K.}~\bibnamefont {Xu}}, \ and\
  \bibinfo {author} {\bibfnamefont {S.}~\bibnamefont {Fu}},\ }\href {\doibase
  https://doi.org/10.1016/j.jcp.2010.05.019} {\bibfield  {journal} {\bibinfo
  {journal} {Journal of Computational Physics}\ }\textbf {\bibinfo {volume}
  {229}},\ \bibinfo {pages} {6715} (\bibinfo {year} {2010})}\BibitemShut
  {NoStop}%
\bibitem [{\citenamefont {Luo}\ and\ \citenamefont {Xu}(2013)}]{luo2013high}%
  \BibitemOpen
  \bibfield  {author} {\bibinfo {author} {\bibfnamefont {J.}~\bibnamefont
  {Luo}}\ and\ \bibinfo {author} {\bibfnamefont {K.}~\bibnamefont {Xu}},\
  }\href {\doibase 10.1007/s11431-013-5334-y} {\bibfield  {journal} {\bibinfo
  {journal} {Science China Technological Sciences}\ }\textbf {\bibinfo {volume}
  {56}},\ \bibinfo {pages} {2370} (\bibinfo {year} {2013})}\BibitemShut
  {NoStop}%
\bibitem [{\citenamefont {Zhou}\ \emph {et~al.}(2017)\citenamefont {Zhou},
  \citenamefont {Xu},\ and\ \citenamefont {Liu}}]{zhou2017simplification}%
  \BibitemOpen
  \bibfield  {author} {\bibinfo {author} {\bibfnamefont {G.}~\bibnamefont
  {Zhou}}, \bibinfo {author} {\bibfnamefont {K.}~\bibnamefont {Xu}}, \ and\
  \bibinfo {author} {\bibfnamefont {F.}~\bibnamefont {Liu}},\ }\href {\doibase
  https://doi.org/10.1016/j.jcp.2017.03.023} {\bibfield  {journal} {\bibinfo
  {journal} {Journal of Computational Physics}\ }\textbf {\bibinfo {volume}
  {339}},\ \bibinfo {pages} {146} (\bibinfo {year} {2017})}\BibitemShut
  {NoStop}%
\bibitem [{\citenamefont {Pan}\ \emph {et~al.}(2016{\natexlab{b}})\citenamefont
  {Pan}, \citenamefont {Xu}, \citenamefont {Li},\ and\ \citenamefont
  {Li}}]{pan2016efficient}%
  \BibitemOpen
  \bibfield  {author} {\bibinfo {author} {\bibfnamefont {L.}~\bibnamefont
  {Pan}}, \bibinfo {author} {\bibfnamefont {K.}~\bibnamefont {Xu}}, \bibinfo
  {author} {\bibfnamefont {Q.}~\bibnamefont {Li}}, \ and\ \bibinfo {author}
  {\bibfnamefont {J.}~\bibnamefont {Li}},\ }\href {\doibase
  https://doi.org/10.1016/j.jcp.2016.08.054} {\bibfield  {journal} {\bibinfo
  {journal} {Journal of Computational Physics}\ }\textbf {\bibinfo {volume}
  {326}},\ \bibinfo {pages} {197} (\bibinfo {year}
  {2016}{\natexlab{b}})}\BibitemShut {NoStop}%
\bibitem [{\citenamefont {Pan}\ and\ \citenamefont
  {Xu}(2015{\natexlab{a}})}]{pan2015third}%
  \BibitemOpen
  \bibfield  {author} {\bibinfo {author} {\bibfnamefont {L.}~\bibnamefont
  {Pan}}\ and\ \bibinfo {author} {\bibfnamefont {K.}~\bibnamefont {Xu}},\
  }\href {\doibase https://doi.org/10.1016/j.compfluid.2015.07.006} {\bibfield
  {journal} {\bibinfo  {journal} {Computers \& Fluids}\ }\textbf {\bibinfo
  {volume} {119}},\ \bibinfo {pages} {250} (\bibinfo {year}
  {2015}{\natexlab{a}})}\BibitemShut {NoStop}%
\bibitem [{\citenamefont {Pan}\ and\ \citenamefont
  {Xu}(2015{\natexlab{b}})}]{pan2015compact}%
  \BibitemOpen
  \bibfield  {author} {\bibinfo {author} {\bibfnamefont {L.}~\bibnamefont
  {Pan}}\ and\ \bibinfo {author} {\bibfnamefont {K.}~\bibnamefont {Xu}},\
  }\href {\doibase 10.4208/cicp.141214.140715s} {\bibfield  {journal} {\bibinfo
   {journal} {Communications in Computational Physics}\ }\textbf {\bibinfo
  {volume} {18}},\ \bibinfo {pages} {985} (\bibinfo {year}
  {2015}{\natexlab{b}})}\BibitemShut {NoStop}%
\bibitem [{\citenamefont {Pan}\ and\ \citenamefont {Xu}(2016)}]{pan2016third}%
  \BibitemOpen
  \bibfield  {author} {\bibinfo {author} {\bibfnamefont {L.}~\bibnamefont
  {Pan}}\ and\ \bibinfo {author} {\bibfnamefont {K.}~\bibnamefont {Xu}},\
  }\href {\doibase https://doi.org/10.1016/j.jcp.2016.05.012} {\bibfield
  {journal} {\bibinfo  {journal} {Journal of Computational Physics}\ }\textbf
  {\bibinfo {volume} {318}},\ \bibinfo {pages} {327} (\bibinfo {year}
  {2016})}\BibitemShut {NoStop}%
\bibitem [{\citenamefont {Wang}\ \emph
  {et~al.}(2016{\natexlab{a}})\citenamefont {Wang}, \citenamefont {Ren},\ and\
  \citenamefont {Li}}]{WANG2016863}%
  \BibitemOpen
  \bibfield  {author} {\bibinfo {author} {\bibfnamefont {Q.}~\bibnamefont
  {Wang}}, \bibinfo {author} {\bibfnamefont {Y.-X.}\ \bibnamefont {Ren}}, \
  and\ \bibinfo {author} {\bibfnamefont {W.}~\bibnamefont {Li}},\ }\href
  {\doibase https://doi.org/10.1016/j.jcp.2016.01.036} {\bibfield  {journal}
  {\bibinfo  {journal} {Journal of Computational Physics}\ }\textbf {\bibinfo
  {volume} {314}},\ \bibinfo {pages} {863 } (\bibinfo {year}
  {2016}{\natexlab{a}})}\BibitemShut {NoStop}%
\bibitem [{\citenamefont {Wang}\ \emph
  {et~al.}(2016{\natexlab{b}})\citenamefont {Wang}, \citenamefont {Ren},\ and\
  \citenamefont {Li}}]{WANG2016883}%
  \BibitemOpen
  \bibfield  {author} {\bibinfo {author} {\bibfnamefont {Q.}~\bibnamefont
  {Wang}}, \bibinfo {author} {\bibfnamefont {Y.-X.}\ \bibnamefont {Ren}}, \
  and\ \bibinfo {author} {\bibfnamefont {W.}~\bibnamefont {Li}},\ }\href
  {\doibase https://doi.org/10.1016/j.jcp.2016.03.048} {\bibfield  {journal}
  {\bibinfo  {journal} {Journal of Computational Physics}\ }\textbf {\bibinfo
  {volume} {314}},\ \bibinfo {pages} {883 } (\bibinfo {year}
  {2016}{\natexlab{b}})}\BibitemShut {NoStop}%
\bibitem [{\citenamefont {Saad}\ and\ \citenamefont
  {Schultz}(1986)}]{saad1986gmres}%
  \BibitemOpen
  \bibfield  {author} {\bibinfo {author} {\bibfnamefont {Y.}~\bibnamefont
  {Saad}}\ and\ \bibinfo {author} {\bibfnamefont {M.~H.}\ \bibnamefont
  {Schultz}},\ }\href@noop {} {\bibfield  {journal} {\bibinfo  {journal} {SIAM
  Journal on Scientific and Statistical Computing}\ }\textbf {\bibinfo {volume}
  {7}},\ \bibinfo {pages} {856} (\bibinfo {year} {1986})}\BibitemShut {NoStop}%
\bibitem [{\citenamefont {Saad}(2003)}]{saad2003iterative}%
  \BibitemOpen
  \bibfield  {author} {\bibinfo {author} {\bibfnamefont {Y.}~\bibnamefont
  {Saad}},\ }\href@noop {} {\emph {\bibinfo {title} {Iterative methods for
  sparse linear systems}}}\ (\bibinfo  {publisher} {SIAM},\ \bibinfo {year}
  {2003})\BibitemShut {NoStop}%
\bibitem [{\citenamefont {Cuthill}\ and\ \citenamefont
  {McKee}(1969)}]{cuthill1969reducing}%
  \BibitemOpen
  \bibfield  {author} {\bibinfo {author} {\bibfnamefont {E.}~\bibnamefont
  {Cuthill}}\ and\ \bibinfo {author} {\bibfnamefont {J.}~\bibnamefont
  {McKee}},\ }in\ \href@noop {} {\emph {\bibinfo {booktitle} {Proceedings of
  the 1969 24th National Conference}}}\ (\bibinfo {organization} {ACM},\
  \bibinfo {year} {1969})\ pp.\ \bibinfo {pages} {157--172}\BibitemShut
  {NoStop}%
\bibitem [{\citenamefont {Meijerink}\ and\ \citenamefont {van~der
  Vorst}(1981)}]{meijerink1981guidelines}%
  \BibitemOpen
  \bibfield  {author} {\bibinfo {author} {\bibfnamefont {J.}~\bibnamefont
  {Meijerink}}\ and\ \bibinfo {author} {\bibfnamefont {H.}~\bibnamefont
  {van~der Vorst}},\ }\href {\doibase
  https://doi.org/10.1016/0021-9991(81)90041-3} {\bibfield  {journal} {\bibinfo
   {journal} {Journal of Computational Physics}\ }\textbf {\bibinfo {volume}
  {44}},\ \bibinfo {pages} {134} (\bibinfo {year} {1981})}\BibitemShut
  {NoStop}%
\bibitem [{\citenamefont {Hackbusch}(1994)}]{hackbusch1994iterative}%
  \BibitemOpen
  \bibfield  {author} {\bibinfo {author} {\bibfnamefont {W.}~\bibnamefont
  {Hackbusch}},\ }\href@noop {} {\emph {\bibinfo {title} {Iterative solution of
  large sparse systems of equations}}},\ Vol.~\bibinfo {volume} {40}\ (\bibinfo
   {publisher} {Springer},\ \bibinfo {year} {1994})\BibitemShut {NoStop}%
\bibitem [{\citenamefont {Liu}\ and\ \citenamefont
  {Zhang}(2017)}]{liu2017accuracy}%
  \BibitemOpen
  \bibfield  {author} {\bibinfo {author} {\bibfnamefont {Y.}~\bibnamefont
  {Liu}}\ and\ \bibinfo {author} {\bibfnamefont {W.}~\bibnamefont {Zhang}},\
  }\href {\doibase https://doi.org/10.1016/j.compfluid.2017.03.008} {\bibfield
  {journal} {\bibinfo  {journal} {Computers \& Fluids}\ }\textbf {\bibinfo
  {volume} {149}},\ \bibinfo {pages} {88} (\bibinfo {year} {2017})}\BibitemShut
  {NoStop}%
\bibitem [{\citenamefont {Kogan}(1969)}]{GeneralSolution-Kogan}%
  \BibitemOpen
  \bibfield  {author} {\bibinfo {author} {\bibfnamefont {M.~N.}\ \bibnamefont
  {Kogan}},\ }\href@noop {} {\emph {\bibinfo {title} {Rarefied Gas Dynamics}}}\
  (\bibinfo  {publisher} {Plenum Press},\ \bibinfo {year} {1969})\BibitemShut
  {NoStop}%
\bibitem [{\citenamefont {Gustafsson}(1975)}]{gustafsson1975convergence}%
  \BibitemOpen
  \bibfield  {author} {\bibinfo {author} {\bibfnamefont {B.}~\bibnamefont
  {Gustafsson}},\ }\href@noop {} {\bibfield  {journal} {\bibinfo  {journal}
  {Mathematics of Computation}\ }\textbf {\bibinfo {volume} {29}},\ \bibinfo
  {pages} {396} (\bibinfo {year} {1975})}\BibitemShut {NoStop}%
\bibitem [{\citenamefont {Li}\ and\ \citenamefont {Ren}(2012)}]{li2012multi}%
  \BibitemOpen
  \bibfield  {author} {\bibinfo {author} {\bibfnamefont {W.}~\bibnamefont
  {Li}}\ and\ \bibinfo {author} {\bibfnamefont {Y.-X.}\ \bibnamefont {Ren}},\
  }\href {\doibase https://doi.org/10.1016/j.jcp.2012.01.029} {\bibfield
  {journal} {\bibinfo  {journal} {Journal of Computational Physics}\ }\textbf
  {\bibinfo {volume} {231}},\ \bibinfo {pages} {4053} (\bibinfo {year}
  {2012})}\BibitemShut {NoStop}%
\bibitem [{\citenamefont {Tam}\ and\ \citenamefont
  {Webb}(1993)}]{tam1993dispersion}%
  \BibitemOpen
  \bibfield  {author} {\bibinfo {author} {\bibfnamefont {C.~K.}\ \bibnamefont
  {Tam}}\ and\ \bibinfo {author} {\bibfnamefont {J.~C.}\ \bibnamefont {Webb}},\
  }\href@noop {} {\bibfield  {journal} {\bibinfo  {journal} {Journal of
  Computational Physics}\ }\textbf {\bibinfo {volume} {107}},\ \bibinfo {pages}
  {262} (\bibinfo {year} {1993})}\BibitemShut {NoStop}%
\bibitem [{\citenamefont {Shu}\ and\ \citenamefont
  {Osher}(1988)}]{shu1988efficient}%
  \BibitemOpen
  \bibfield  {author} {\bibinfo {author} {\bibfnamefont {C.-W.}\ \bibnamefont
  {Shu}}\ and\ \bibinfo {author} {\bibfnamefont {S.}~\bibnamefont {Osher}},\
  }\href {\doibase https://doi.org/10.1016/0021-9991(88)90177-5} {\bibfield
  {journal} {\bibinfo  {journal} {Journal of Computational Physics}\ }\textbf
  {\bibinfo {volume} {77}},\ \bibinfo {pages} {439} (\bibinfo {year}
  {1988})}\BibitemShut {NoStop}%
\bibitem [{\citenamefont {Wang}\ and\ \citenamefont
  {Ren}(2015)}]{wang2015accurate}%
  \BibitemOpen
  \bibfield  {author} {\bibinfo {author} {\bibfnamefont {Q.}~\bibnamefont
  {Wang}}\ and\ \bibinfo {author} {\bibfnamefont {Y.-X.}\ \bibnamefont {Ren}},\
  }\href@noop {} {\bibfield  {journal} {\bibinfo  {journal} {Journal of
  Computational Physics}\ }\textbf {\bibinfo {volume} {284}},\ \bibinfo {pages}
  {648} (\bibinfo {year} {2015})}\BibitemShut {NoStop}%
\bibitem [{\citenamefont {Ghia}\ \emph {et~al.}(1982)\citenamefont {Ghia},
  \citenamefont {Ghia},\ and\ \citenamefont {Shin}}]{ghia1982high}%
  \BibitemOpen
  \bibfield  {author} {\bibinfo {author} {\bibfnamefont {U.}~\bibnamefont
  {Ghia}}, \bibinfo {author} {\bibfnamefont {K.}~\bibnamefont {Ghia}}, \ and\
  \bibinfo {author} {\bibfnamefont {C.}~\bibnamefont {Shin}},\ }\href {\doibase
  https://doi.org/10.1016/0021-9991(82)90058-4} {\bibfield  {journal} {\bibinfo
   {journal} {Journal of Computational Physics}\ }\textbf {\bibinfo {volume}
  {48}},\ \bibinfo {pages} {387} (\bibinfo {year} {1982})}\BibitemShut
  {NoStop}%
\end{thebibliography}%
\clearpage
\listoffigures
\clearpage

\end{document}